\documentclass[prb,twocolumn,showpacs,floatfix]{revtex4}
\usepackage{graphicx}
\usepackage{amssymb}
\usepackage{amsmath}
\usepackage{xspace}
\usepackage{url}

\begin{document} %%%%%%%%%%%%%%%%%%%%%%%%%%%%%%%%%%%%%%%%%%%%%%%%%%%%%%%%%%
%------------------------------------------------------------------------------ 
% Title
%------------------------------------------------------------------------------
%\title{Phonon-mediated desorption of electrons}
\title{Phonon-mediated desorption of image-bound electrons from dielectric surfaces}
%\title{Phonon-mediated desorption of image-bound electrons}
%\title{Phonon-mediated desorption of electrons bound in polarization-induced surface states}
%------------------------------------------------------------------------------ 
% Authors
%------------------------------------------------------------------------------
%------------------------------------------------------------------------------ 
% Date
%------------------------------------------------------------------------------
\author{R. L. Heinisch, F. X. Bronold, and 
H. Fehske}
\affiliation{Institut f{\"ur} Physik,
             Ernst-Moritz-Arndt-Universit{\"a}t Greifswald,
             17489 Greifswald,
             Germany}

\date{\today}
\begin{abstract}
A complete kinetic modeling of an ionized gas in contact with a surface requires the knowledge of the 
electron desorption time and the electron sticking coefficient. We calculate the desorption time for phonon-mediated
desorption of an image-bound electron, as it occurs, for instance, on dielectric surfaces where desorption 
channels involving internal electronic degrees of freedom are closed. Because of the large depth of the
polarization-induced surface potential with respect to the Debye energy multi-phonon processes are important. 
To obtain the desorption time, we use a quantum-kinetic rate equation for the occupancies of the bound 
electron surface states, taking two-phonon processes into account in cases where one-phonon processes yield
a vanishing transition probability, as it is sufficient, for instance, for graphite. For an electron desorbing 
from a graphite surface at $360~K$ we find a desorption time of $2\cdot 10^{-5}~s$.
We also demonstrate that depending on the potential depth and bound state level spacing the 
desorption scenario changes. In particular, we show that desorption via cascades over bound states 
dominates unless direct one-phonon transitions from the lowest bound state to the continuum are 
possible.
\end{abstract}
\maketitle

\section{Introduction}
\label{Introduction}

% \subsection{Motivation}

Whenever at the surface of a solid the vacuum level falls inside an energy gap, that is, whenever 
the electron affinity of the surface is negative, polarization-induced external surface states 
(image states) exist, as it is known from macroscopic electrodynamics~\cite{Boettcher52}. Originally
predicted~\cite{CC69} for the surfaces of liquid and solid He, Ne, ${\rm H_2}$, and ${\rm D_2}$ 
the existence of image states has by now been experimentally verified for a great number of 
metallic~\cite{DAG84,SH84,WHJ85,JDK86,Fauster94,HSR97,Hoefer99,FW05,GSB08}
as well as insulating~\cite{Cole74,LMT99,KWR07} surfaces. In addition, there exist a variety of 
dielectric materials, for instance, diamond~\cite{HKV79,CRL98,YMN09}, boron nitride~\cite{LSG99},
and alkali-earth metal oxides~\cite{RWK03,BKP07,MS08}, which have surfaces with a negative electron 
affinity. They should thus support image states. Interesting in this respect are also 
electro-negative dielectric structures used in electron emitting devices, such as, cesium-doped 
silicon oxide films~\cite{DMS99,GDK05,MCK06} and GaAs-based heterostructures~\cite{BGY05,BGP05,YYK07}.

In contrast to intrinsic surface states~\cite{Spanjaard96}, originating either from the sudden
disappearance of the periodic lattice potential or unsaturated bonds at the surface, image states 
are not localized at the edge but typically a few $\AA$ in front of the solid. An external 
electron approaching the solid from the vacuum with a kinetic energy below the lowest unoccupied
intrinsic electron state of the surface may thus get trapped (adsorbed) in these states provided
it can get rid of its excess energy. Once it is trapped it may de-trap again (desorb) if it 
gains enough energy from the solid. Hence, in addition to elastic and inelastic scattering, 
the interaction of low-energy electrons with surfaces may encompass physisorption --
the polarization-induced temporary binding of an electron to the surface. 

Unlike physisorption of neutral atoms and molecules, which has been studied in great detail 
ever since the seminal works of Lennard-Jones and
collaborators~\cite{LJS35,Strachan35,LJD36,BY73,Brenig82,GKS80,GKT80a,GKT80b,KG86,Brenig87}, 
physisorption of electrons has been hardly investigated. It is only until recently that we pointed 
out~\cite{BFKD08,BDF09} that the charging of surfaces in contact with an ionized gas, as it
occurs, for instance, in the interstellar medium~\cite{Whipple81,DS87,Mann08}, in the upper 
atmosphere~\cite{RL01}, in dusty laboratory plasmas~\cite{FIK05,Ishihara07}, and in dielectrically 
bounded low-temperature plasmas~\cite{GMB02,Kogelschatz03,KCO04,SAB06,SLP07,LLZ08}, could 
be perhaps microscopically understood as an electronic physisorption process. 

Parameters characterizing physisorption of electrons at surfaces are the electron sticking 
coefficient $s_{\rm e}$ and the electron desorption time $\tau_{\rm e}$. Little is quantitatively known
about these parameters, although they are rather important for a complete kinetic description of bounded 
gas discharges (as it is in fact also the case for the sticking coefficient and desorption time 
of neutral particles which play a central role for the kinetic modeling of bounded neutral 
gases~\cite{Kuscer78,FM09a,FM09b}).  
Very often
%, $s_{\rm e}\approx 0.1-1$ 
%and $\tau_{\rm e}^{-1}\approx 0$ without further justification. Sometimes 
$s_{\rm e}$ and $\tau_{\rm e}$ are simply used as adjustable parameters. 

In view of the importance of $s_{\rm e}$ and $\tau_{\rm e}$ for bounded plasmas, we 
adopted in Ref.~\cite{BDF09} the quantum-kinetic approach originally developed for the theoretical 
description of physisorption of neutral particles~\cite{KG86,Brenig87} to calculate $s_{\rm e}$ and 
$\tau_{\rm e}$ for a metallic surface. Neglecting crystal-induced surface states and describing 
the metal within the jellium model we obtained for an ideal surface with a classical image potential
$s_{\rm e}\approx 10^{-4}$ and $\tau_{\rm e}\approx 10^{-2}~s$. Although $s_{\rm e}$ seems 
to be rather small, the product $s_{\rm e}\tau_{\rm e}\approx 10^{-6}~s$, which is the order of 
magnitude we expected from our study of charging of dust particles in low-temperature 
plasmas~\cite{BFKD08}.

Physisorption of an external electron implies energy exchange between the electron and the 
electronic and/or vibrational elementary excitations of the surface. For a metallic surface  
creation and annihilation of internal electron-hole pairs seem to be the main reason for electron energy 
relaxation at the surface~\cite{NNS86,WJS92}. For a dielectric surface, however, the typical energy of 
an internal electron-hole pair is of the order of the energy gap, that is, for the dielectrics 
we are interested in, a few electron volts. For typical surface temperatures this is way 
too large for electron-hole pairs to cause energy relaxation at the surface. At dielectric 
surfaces it has to be rather the creation and annihilation of phonons which leads to 
electron energy relaxation.

For dielectrics with a large dielectric constant and a large energy gap, the level 
spacing of the two lowest states in the (polarization-induced) surface potential turns out to 
exceed the maximum phonon energy, which is, within the Debye model, the Debye energy. Hence, 
in contrast to physisorption of neutral particles, which typically involves a few bound states 
with energy spacings not exceeding the Debye energy~\cite{KG86}, physisorption of electrons at 
(this type of) dielectric surfaces takes place in a deep potential supporting deep bound states 
whose energy spacings may be larger than the Debye energy. Relaxation channels involving 
internal electronic degrees of freedom being closed, because of the large gap, electron energy 
relaxation, and hence sticking and desorption of electrons, has to be controlled by
multi-phonon processes. 

Typical dielectric plasma boundaries are, in dusty plasmas~\cite{FIK05,Ishihara07}, 
graphite and melamine-formaldehyde, and in dielectric barrier 
discharges~\cite{GMB02,Kogelschatz03,KCO04,SAB06,SLP07,LLZ08} Duran glass, silicon dioxide,
and aluminum oxide. We suspect on empirical grounds that plasma boundaries 
always support image states, if not intrinsically then due to chemical 
contamination from the discharge. Based on this hypothesis we investigate in the following, 
employing a simple model for the polarization-induced interaction between an electron and 
a dielectric surface~\cite{RM72,EM73}, the desorption of an image-bound electron from a 
dielectric surface. We are particularly interested in how 
%for varying potential depths
multi-phonon processes affect the competition between direct desorption, that is, 
the direct transition between bound and unbound surface states, and cascading 
desorption~\cite{GKT80a}, that is, the successive climbing up of the ladder of bound surface 
states until the continuum is reached.

For the plasma boundaries just mentioned, image states have been so far only observed 
for graphite~\cite{LMT99} (see Table \ref{materialtable} for the relevant material 
parameters). Surprisingly, the measured binding energy of the lowest image state, 
\(E^{\rm exp}_1\approx -0.85~eV\), is lower than the energy of the lowest 
bound state in the classical image potential, which should be in fact a lower bound~\cite{CC69}. 
Indeed, for \(\epsilon_s=13.5\), the dielectric constant of graphite, \(E_1^{\rm cl}\approx -0.63~eV\). 
Taking either \(E^{\rm exp}_1\) or \(E^{\rm cl}_1\) in conjunction with 
\(\hbar\omega_D\approx 0.22~eV\), the Debye energy for graphite, \(3.9\) 
or \(2.9\) phonons would be required for a direct transition to the continuum. 
The probability for an electron to de-trap from the lowest image state of graphite via such a 
transition would be accordingly small. Cascades using higher lying bound states as intermediaries
should therefore become rather important. The largest energy difference to overcome would 
then be the one between the two lowest bound states. For the classical image potential, e.g., 
this difference is $0.47~eV$, implying that at most $2.35$ phonons are required for getting
a cascade running from the lowest level.

From these estimates we conclude that for graphite, with its rather high Debye energy, 
the number of phonons involved in physisorption of electrons is small enough to use it as 
an expansion parameter for the transition probability. Taking moreover the recoil energy into account 
the dipole-active elementary excitation responsible for the polarization-induced surface potential 
imparts onto the external electron~\cite{EM73}, two-phonon processes even turn out to suffice.
The other dielectrics have a much smaller Debye energy. 
The number of phonons involved is thus much larger. Instead of a brute force expansion other approaches 
seem to be more suitable in these cases~\cite{Gumhalter96}.

The outline of the remaining paper is as follows. 
%The next three sections explain our formalism. 
First, in Section II, we set up the quantum-kinetic rate equation for the occupancies
of bound surface states and introduce a classification scheme for the depth of the 
surface potential. In Section III we describe the microscopic model for the electron-surface 
interaction, including the static part which provides the surface states 
involved in physisorption and the dynamic part which drives the transitions between these 
states and is thus responsible for desorption. In Section IV we calculate the transition 
probability up to fourth order in the displacement field, thereby taking one- and two-phonon processes 
into account, which we believe to be sufficient for graphite. This calculation is very
lengthy~\cite{Rafael09} and cannot be totally reproduced here. Three appendices provide the 
required mathematical details. Finally, in Section V we present and discuss our results 
before we conclude in Section VI. 

\section{Desorption from many bound states}
\label{Desorption from many bound states}

Following Gortel, Kreuzer and Teshima~\cite{GKT80a}, the kinetics in a manifold of bound surface 
states can be described by a quantum-kinetic rate equation. Assuming that once the electron is in an 
unbound surface state it is immediately pushed away from the surface, which is reasonable if we consider 
the electron as a test electron desorbing from a negatively charged surface (see Ref.~\cite{BDF09} for
more details), the time evolution of the occupancies of the bound surface states for the (test) electron 
is given by~\cite{GKT80a}
\begin{align}
\frac{\mathrm{d}}{\mathrm{d}t} n_{q}(t) = & \sum_{q'} \left[W_{q q'} n_{q'}(t) - W_{q' q} n_{q}(t)\right] 
-W_{cq}n_{q}(t)\text{ ,}
\label{desrateeq}
\end{align}
where \(W_{qq'}\) is the probability for a transition from state \(q^\prime\) to state \(q\) and 
\(W_{cq}=\sum_k W_{kq}\) is the probability for a 
transition from the bound state \(q\) to the continuum. In compact matrix notation (\ref{desrateeq}) 
may be rewritten as 
\begin{align}
\frac{\mathrm{d}}{\mathrm{d}t}\mathbf{n}=\mathbf{T}\mathbf{n} \text{ ,}
\end{align}
where \(\mathbf{n}\) is the \(N\)-dimensional column vector of the occupancies of the bound surface states
and \(\mathbf{T}\) is the matrix of the transition probabilities.

To determine the formal solution of this equation,
\begin{align}
\mathbf{n}(t)=\exp[\mathbf{T}t]\mathbf{n}(0) \text{ ,}
\end{align}
the eigenvalue equation for the matrix \(\mathbf{T}\) has to be solved. In general, \(\mathbf{T}\) is not 
symmetric. Thus, there are right and left eigenvectors~\cite{GKT80a}, 
\(\mathbf{e}^{\kappa}\) and \(\mathbf{\tilde{ e}}^{\kappa}\), respectively, 
%\begin{align}
%&\mathbf{T}\mathbf{e}^{(\kappa)}=-\lambda_\kappa \mathbf{e}^{(\kappa)} \text{ ,}\\
%& \mathbf{\tilde e}^{(\kappa)} \mathbf{T} = -\lambda_\kappa  \mathbf{\tilde e}^{(\kappa)}\text{ ,}
%\end{align}
which can be chosen to be orthogonal to each other. 
%to satisfy
%\begin{align}
%\sum_i {\tilde{\rm e}}_i^{(\kappa)}{\rm e}_i^{(\kappa^\prime)} = \delta_{\kappa,\kappa^\prime}\text{ .} 
%\end{align}
In terms of the right eigenvectors of \(\mathbf{T}\), 
%the occupancies at time \(t\) are simply given by 
\begin{align}
\mathbf{n}(t)=\sum_\kappa f^\kappa e^{-\lambda_\kappa t} \mathbf{e}^{\kappa} \text{ ,} \label{gensolrateeq}
\end{align}
where the coefficients \(f^\kappa\) are determined by decomposing the initial distribution into eigenfunctions 
according to
\begin{align}
\mathbf{n}(0)=\sum_\kappa f^\kappa \mathbf{e}^{\kappa}\text{ .} 
\end{align}
Due to the losses to the continuum all eigenvalues \(-\lambda_\kappa\) turn out to be negative~\cite{GKT80a}.
Hence, for sufficiently long times the image-bound electron escapes into the continuum and the bound state
occupation vanishes, i.e. \(n_q (t \rightarrow \infty) = 0 ~\forall q\). 
If the transitions leading to losses to the continuum are much slower than the transitions between bound 
states, i.e. \( W_{cq} \ll W_{qq'} \), the bound electron evaporates slowly into the continuum. One 
eigenvalue, \(\lambda_0\), is then considerably smaller than all the others and its right eigenvector 
corresponds to the equilibrium distribution \(n_q^\mathrm{eq}={\rm e}_q^{(0)}\). The general solution 
(\ref{gensolrateeq}) can then be split into two terms~\cite{GKT80a},
\begin{align}
n_q(t)=f^{0} {\rm e}_q^{0}e^{-\lambda_0 t}+\sum_{\kappa > 0} f^{\kappa} 
{\rm e}_q^{\kappa} e^{-\lambda_\kappa t} \text{ ,}
\end{align}
where the first term gives the time evolution for an equilibrium occupation of the bound states whereas the 
second term describes the fast equilibration of a distortion of the equilibrium occupation at the beginning 
of the desorption process. It is subject to much faster transitions which will be completed soon after the 
beginning of the desorption process. Since the fate of the electron for long times depends only on the 
equilibrium occupation, we identify the inverse of the desorption time with the lowest eigenvalue,
\begin{align}
\tau_{e}^{-1} = \lambda_0 \text{ .}
\end{align}

This conceptual framework of desorption requires surface states and transition probabilities between them as 
input. For dielectric surfaces the transitions are driven by phonons whose energy scale, within the Debye model,
is the Debye energy. It is therefore natural to measure energies in units of the Debye energy 
$\hbar \omega_D=k_BT_D$.  Important parameters characterizing the potential depth are then 
\begin{align}
\epsilon_q=\frac{E_q}{\hbar \omega_D} \quad \text{and}
\quad \Delta_{qq^\prime}=\frac{E_q-E_{q^\prime}}{\hbar \omega_D},
\end{align}
where $E_q<0$ denotes the energy of the $q$th bound state. We call the surface potential
shallow if the lowest bound state is at most one Debye energy beneath the continuum, i.e. \(\epsilon_1>-1\),
one-phonon deep if the energy difference between the lowest two bound states is less than one Debye energy,
i.e. \(\Delta_{12}>-1\), two-phonon deep if the energy difference between the lowest two bound states is 
between one and two Debye energies, i.e. \(-1>\Delta_{12}>-2\). 
 
Shallow and one-phonon deep potentials are typical for physisorption of neutral atoms and 
molecules. Because of the strong polarization-induced interaction between an external electron and a 
surface, physisorption of electrons, however, typically takes place in at least two-phonon
deep surface potentials. Multi-phonon processes should thus play an important role.

\section{Electron-surface interaction}
\label{Electron-surface interaction}
%\subsection{Static electron-surface interaction}

An electron in front of a solid surface feels a polarization-induced attraction to the surface because 
of the coupling to dipole-active excitations of the solid. For a dielectric material 
the relevant modes are optical surface phonons~\cite{RM72,EM73}.

If, for a dielectric solid with negative electron affinity, the kinetic energy of an external electron 
is less than the negative of the electron affinity, the electron cannot enter the solid, which, for the 
purpose of the calculation, we assume to fill the whole left half-space (\(z\le 0\)), being terminated
at $z=0$ with a surface whose lateral extension $A$ is eventually made infinitely large. Evans and 
Mills~\cite{EM73} studied this situation by variational means. They found that far from the surface, 
the interaction potential is the classical image potential known from elementary electrostatics but
close to the surface the interaction potential is strongly modified by the recoil energy resulting from 
the momentum transfer parallel to the surface when the electron absorbs or emits a (dipole-active)
surface phonon.

The recoil energy makes the interaction potential not only nonlocal for distances less than the bulk 
polaron radius \(z_s=\sqrt{\hbar/2m\omega_s}\), where \(m\) is the mass of the electron, \(\epsilon_s\) 
is the static dielectric constant, and \(\omega_s=\omega_T\sqrt{(1+\epsilon_s)/2}\) is the frequency of 
the surface phonon (\(\omega_T\) is the TO-phonon frequency). Most importantly, it makes 
the interaction potential finite at the surface, in contrast to the singular behavior of the classical 
image potential. Denoting the lateral two-dimensional momentum transfer by \(\vec{K} \), the simplest 
regular local approximation to the true interaction potential is~\cite{EM73}
\begin{align}
V(z)=-e^2\frac{\epsilon_s-1}{\epsilon_s+1}\frac{\pi}{A}\sum_{\vec{K}}
\frac{1}{K}\frac{e^{-2K  |z|}}{1+\frac{\hbar}{2m\omega_s} K^2} 
\label{varapprox} \text{ ,}
\end{align}
where $K$ is the magnitude of the vector $\vec{K}$.
It can be considered as a dynamically corrected classical image potential. Indeed, neglecting in the 
denominator the recoil energy, \(\hbar K^2/2m \), the integral over \(\vec{K}\) can be 
easily performed and leads to 
\begin{align}
V_{\rm cl}(z) = -\frac{e^2}{4} \frac{\epsilon_s -1}{\epsilon_s+1}\frac{1}{z} \text{ ,} \label{classicalimage}
\end{align}
which is the classical image potential. 

The dynamically corrected image potential (\ref{varapprox}) is attractive. The solution of the 
corresponding Schr\"odinger equation will thus yield bound and unbound surface states. To make an analytical 
solution feasible, we fit the dynamically corrected image potential (\ref{varapprox}) to a \(1/z\) potential 
that is shifted along the \(z\) axis. Forcing the two potentials to coincide at the surface, that is, at 
\(z=0\),  we obtain
\begin{align}
V(z)\approx -\frac{e^2}{4} \frac{\epsilon_s -1}{\epsilon_s+1}\frac{1}{z+z_c} 
\label{Vdyn}
\end{align}
with \(z_c=z_s/\pi\).
After the transformation \(z\rightarrow z-z_c\) the Schr\"odinger equation corresponding to the shifted 
surface potential reads, in dimensionless variables  \(x=z /a_B\) and \(\eta = 2 \hbar^2 E /m e^4\),  
\begin{align}
\phi^{\prime\prime}(x)+\left(\frac{2 \Lambda_0}{x}+\eta \right) \phi(x)=0 \text{ ,} \label{sesufracestates}
\end{align}
where \(a_B=\hbar^2/me^2\) is the Bohr radius and \(\Lambda_0=(\epsilon_s-1)/4(\epsilon_s+1)\). Assuming
that electrons cannot enter the dielectric surface, we solve Eq. (\ref{sesufracestates}) 
with the boundary condition \(\phi(x_c)=0\) where \(x_c=z_c/a_B\). The wave functions and energies for 
bound and unbound surface states, together with the additional boundary conditions we have to impose on them,
are given in Appendix \ref{schrss}.

%\subsection{Dynamic electron-surface interaction}

Transitions between the eigenstates are due to dynamic perturbations of the surface potential. The surface
potential is very steep near the surface. A strong perturbation arises therefore from the longitudinal 
acoustic phonon perpendicular to the surface which causes the surface plane to oscillate. 

Including this type of surface vibrations and using the eigenstates of (\ref{sesufracestates}) 
as a basis, the Hamiltonian for the surface electron can be split into three parts,
\begin{align}
H=H_{e}^\text{static} + H^0_{ph} + H_{e-ph}^\text{dyn} \text{ ,} \label{elsurffulham}
\end{align}
where the first term is the Hamiltonian for the electron in the static surface potential,
\begin{align}
H_{e}^\text{static}=\sum_{q} E_q c_q^\dagger c_q \text{ ,}
\end{align}
the second term is the Hamiltonian of the free acoustic phonons,
\begin{align}
H^{0}_{ph}=\sum_Q \hbar \omega_Q b_Q^\dagger b_Q \text{ ,} \label{freephononhamilt}
\end{align}
where \( Q \) denotes a one-dimensional perpendicular wave vector, and the last term is the dynamic perturbation due 
to surface vibrations. Denoting for simplicity both bound and unbound eigenstates of the surface potential by \(|q\rangle\), 
it is given by
\begin{align}
H_{e-ph}^\text{dyn}=\sum_{q,q^\prime} \langle q^\prime | V_p (u,z) | q \rangle c_{q^\prime}^\dagger c_q \text{ .} 
\label{vibperturb}
\end{align}
The displacement of the surface \(u\) is related to the phonon creation and annihilation operators in the
usual way, 
\begin{align}
u=\sum_Q \sqrt{\frac{\hbar}{2\mu \omega_Q N_s}} (b_Q+b_{-Q}^\dagger) \text{  }
\end{align}
with \(\mu\) the mass of the unit cell of the lattice. The perturbation \(V_p(u,z)\) can be identified as the 
difference between the displaced shifted surface potential and the static shifted surface potential.
Recalling (\ref{Vdyn}) and the transformation \(z\rightarrow z-z_c\), it reads 
\begin{align}
V_p(z,u)=-\frac{e^2\Lambda_0}{z+u}+\frac{e^2 \Lambda_0}{z} \text{ ,}
\label{fullperturb}
\end{align}
which, gearing towards a multi-phonon calculation~\cite{GKT80a}, we expand in a Taylor series in \(u\), 
\begin{align}
V_p(z,u)=\frac{e^2 \Lambda_0}{z^2}u-\frac{e^2 \Lambda_0}{z^3}u^2+\frac{e^2\Lambda_0}{z^4}u^3+O\left(u^{4}\right)~. 
\label{uexpanpotential}
\end{align}

\section{Transition Probabilities}
\label{Calculation of the rates}
\subsection{Preparatory considerations}

We intend to calculate the desorption time taking one- and two-phonon processes into account. Hence, we need to evaluate 
the transition probabilities \(W_{qq'}\) for one- and two-phonon processes. In general, multi-phonon processes have 
two possible 
origins~\cite{BY73}: (i) multi-phonon terms in the perturbation of the surface potential (\ref{uexpanpotential}) and (ii) 
multiple actions of the perturbation as it is encoded in the \(T\)-matrix corresponding to \(H_{e-ph}^{\rm dyn}\).

Using expansion (\ref{uexpanpotential}) the dynamic perturbation \(H_{e-ph}^\text{dyn}\) can be classified by the 
order in \(u\). Up to third order,
\begin{align}
H_{e-ph}^\text{dyn}=V_{1}+V_{2}+V_{3}+O\left(u^{4}\right) \text{ ,}
\end{align}
where in second quantized form
\begin{align}
V_{1}=&\sum_{Q}\sum_{q,q^{\prime}}G_{q,q^{\prime}}^{1}\left(Q\right)\left(\mathrm{b}_{Q}+\mathrm{b}_{-Q}^{\dagger}\right) 
c_{q}^{\dagger}c_{q^{\prime}} \text{ ,} \label{vau1} \\
V_{2}=&-\sum_{Q_{1},Q_{2}}\sum_{q,q^{\prime}}G_{q,q^{\prime}}^{2}\left(Q_{1},Q_{2}\right) 
\left(\mathrm{b}_{Q_{1}}+\mathrm{b}_{-Q_{1}}^{\dagger}\right) \nonumber \\
&\quad \times \left(\mathrm{b}_{Q_{2}}+\mathrm{b}_{-Q_{2}}^{\dagger}\right)  c_{q}^{\dagger}c_{q^{\prime}} \text{ ,}  
\label{vau2}\\
V_{3}=&\sum_{Q_{1},Q_{2},Q_{3}}\sum_{q,q^{\prime}} G_{q,q^{\prime}}^{3}\left(Q_{1},Q_{2},Q_{3} \right) 
\left(\mathrm{b}_{Q_{1}}+\mathrm{b}_{-Q_{1}}^{\dagger}\right) \nonumber \\
& \quad \times \left(\mathrm{b}_{Q_{2}}+\mathrm{b}_{-Q_{2}}^{\dagger}\right) 
\left(\mathrm{b}_{Q_{3}}+\mathrm{b}_{-Q_{3}}^{\dagger}\right) c_{q}^{\dagger}c_{q^{\prime}} \label{vau3}~.
\end{align}
The matrix element of the electron-phonon interaction,
\begin{align}
G_{q,q^{\prime}}^{n}\left(Q_{1},\dots ,Q_{n} \right)=\left(\frac{\hbar}{2\mu N_s} \right)^{n/2}\frac{e^2 
\Lambda_0  Z_{q,q^\prime}^{n+1}}{\sqrt{\omega_{Q_1}\dots\omega_{Q_n}}} \text{ ,}
\end{align}
involves the electronic matrix element,
\begin{align}
Z_{q,q^{\prime}}^{n}=\langle q|\frac{1}{z^{n}}|q^{\prime}\rangle \text{ ,} 
\end{align}
whose evaluation is sketched in Appendix \ref{Approximations to the wave functions and matrix elements}.

Quite generally, the transition probability from an electronic state \(|q\rangle\) to an electronic 
state \(|q^\prime\rangle\) is given by~\cite{BY73}
\begin{align}
\mathcal{R}\left(q^\prime,q\right)=&\frac{2\pi}{\hbar}\sum_{s,s^{\prime}}
\frac{e^{-\beta E_{s}}}{\sum_{s^{\prime \prime}}e^{-\beta E_{s^{\prime \prime}}}} 
\left| \langle s^{\prime},q^{\prime}|T|s,q \rangle \right|^{2} \nonumber \\
&\times \delta \left(E_{s}-E_{s^{\prime}}+E_{q}-E_{q^{\prime}} \right) \text{ ,} \label{scatteringR}
\end{align}
where \(\beta=(k_BT_s)^{-1}\) with \(T_s\) the surface temperature; \(|s\rangle\) and \(|s^\prime \rangle\) 
are initial and final phonon states. We are only interested in the transition between electronic 
states. It is thus natural to average in (\ref{scatteringR}) over all phonon states. The delta function guarantees 
energy conservation. 

The \(T\)-matrix describing the interaction between the external electron and the acoustic phonons 
obeys the operator equation,
\begin{align}
T=H_{e-ph}^\text{dyn}+H_{e-ph}^\text{dyn}GH_{e-ph}^\text{dyn}\text{ ,}
\end{align}
where \(G\) satisfies,
\begin{align}
G=G_0+G_0H_{e-ph}^\text{dyn}G \text{ ,}
\end{align}
and \(G_0\) is given by
\begin{align}
G_0=(E-H_0+i\epsilon)^{-1} \text{  } \label{elphresolvent}
\end{align}
with \(H_0=H_{e}^\mathrm{static}+ H^0_{ph}\).

For a two-phonon process we
\( \left| \langle s^{\prime},q^{\prime}|T|s,q \rangle \right|^{2}\) 
in fourth order in \(u\). We thus iterate \(T\) up to third order in \(u\),
\begin{align}
T=&V_{1}+V_{2}+V_{3}+V_{1}G_{0}V_{1}+V_{1}G_{0}V_{2} 
\nonumber\\ 
&+V_{2}G_{0}V_{1}+V_{1}G_{0}V_{1}G_{0}V_{1}+\mathcal{O}\left(u^4\right), 
\label{tmatiteration}
\end{align}
and write for the transition probability (\ref{scatteringR})
\begin{align}
\mathcal{R}(q^\prime,q)=\sum_{n=1}^{17} \mathcal{R}^n (q^\prime,q) \text{ ,}
\end{align}
where the individual transition probabilities \(\mathcal{R}^n (q^\prime,q)\) can be classified by 
their order in \(u\). 
The term of \(\mathcal{O}(u^{2}) \),
\begin{align} \mathcal{R}^{1} : 
\langle s^{\prime},q^{\prime}|V_{1}|s,q\rangle \langle s,q|V_{1}^{\ast}|s^{\prime},q^{\prime}\rangle \text{ ,} 
\label{Grule}
\end{align}
gives rise to the standard golden rule approximation.

\noindent 
Transition probabilities of \(\mathcal{O}(u^{3}) \) vanish as the expectation value of an odd number of phonon 
creation or annihilation operators is zero. We can thus drop from the calculation the terms
\begin{align} &\mathcal{R}^{2} :  \langle s^{\prime},q^{\prime}|V_{1}|s,q\rangle \langle s,q|V_{2}^{\ast}
|s^{\prime},q^{\prime}\rangle \text{ ,}    
\\ & \mathcal{R}^{4} : \langle s^{\prime},q^{\prime}|V_{1}|s,q\rangle \langle s,q|V_{1}^{\ast}G_{0}^{\ast}V_{1}^{\ast}
|s^{\prime},q^{\prime}\rangle \text{ ,}   
\\ &\mathcal{R}^{8} : \langle s^{\prime},q^{\prime}|V_{2} |s,q\rangle \langle s,q|V_{1}^{\ast} 
|s^{\prime},q^{\prime}\rangle~, 
\\ &\mathcal{R}^{12} : \langle s^{\prime},q^{\prime}|V_{1}G_{0}V_{1} 
|s,q\rangle\langle s,q|V_{1}^{\ast} |s^{\prime},q^{\prime}\rangle~.   \end{align}

\noindent
The remaining transition probabilities are of \(\mathcal{O}(u^{4}) \) and describe two-phonon processes, 
\begin{align} & \mathcal{R}^{3}  : \langle s^{\prime},q^{\prime}|V_{1}|s,q\rangle \langle s,q|V_{3}^{\ast}|s^{\prime},q^{\prime}\rangle \text{ ,}  \\ &\mathcal{R}^{5} : \langle s^{\prime},q^{\prime}|V_{1} |s,q\rangle\langle s,q| V_{2}^{\ast}G_{0}^{\ast}V_{1}^{\ast}|s^{\prime},q^{\prime}\rangle \text{ ,}  \\ &\mathcal{R}^{6} : \langle s^{\prime},q^{\prime}|V_{1} |s,q\rangle \langle s,q| V_{1}^{\ast}G_{0}^{\ast}V_{2}^{\ast}|s^{\prime},q^{\prime}\rangle \text{ ,}  \\ &\mathcal{R}^{7} : \langle s^{\prime},q^{\prime}|V_{1} |s,q\rangle \langle s,q|V_{1}^{\ast}G_{0}^{\ast}V_{1}^{\ast}G_{0}^{\ast}V_{1}^{\ast} |s^{\prime},q^{\prime}\rangle \text{ ,}   \\ &\mathcal{R}^{9} : \langle s^{\prime},q^{\prime}|V_{2} |s,q\rangle \langle s,q|V_{2}^{\ast} |s^{\prime},q^{\prime}\rangle \text{ ,} \label{rr9} \\ &\mathcal{R}^{10} : \langle s^{\prime},q^{\prime}|V_{2} |s,q\rangle \langle s,q|V_{1}^{\ast}G_{0}^{\ast}V_{1}^{\ast} |s^{\prime},q^{\prime}\rangle \text{ ,} \label{rr10} \\ &\mathcal{R}^{11} : \langle s^{\prime},q^{\prime}|V_{3} |s,q\rangle \langle s,q|V_{1}^{\ast} |s^{\prime},q^{\prime}\rangle \text{ ,}  \\ &\mathcal{R}^{13} : \langle s^{\prime},q^{\prime}|V_{1}G_{0}V_{1} |s,q\rangle\langle s,q|V_{2}^{\ast} |s^{\prime},q^{\prime}\rangle \text{ ,} \label{rr13}  \\ &\mathcal{R}^{14} : \langle s^{\prime},q^{\prime}|V_{1}G_{0}V_{1} |s,q\rangle \langle s,q| V_{1}^{\ast}G_{0}^{\ast}V_{1}^{\ast}|s^{\prime},q^{\prime}\rangle \text{ ,} \label{rr14} \\ &\mathcal{R}^{15} : \langle s^{\prime},q^{\prime}|V_{1}G_{0}V_{2}|s,q\rangle\langle s,q|V_{1}^{\ast}|s^{\prime},q^{\prime}\rangle \text{ ,}  \\ &\mathcal{R}^{16} : \langle s^{\prime},q^{\prime}|V_{2}G_{0}V_{1}|s,q\rangle \langle s,q|V_{1}^{\ast}|s^{\prime},q^{\prime}\rangle~,
\\&\mathcal{R}^{17} : \langle s^{\prime},q^{\prime}|V_{1}G_{0}V_{1}G_{0}V_{1}|s,q\rangle \langle s,q|V_{1}^{\ast}|s^{\prime},q^{\prime}\rangle \text{ .}  \end{align}

A complete two-phonon calculation would take all these transition probabilities into account as they stand. This 
is however not always necessary. In the next subsection we show that the two-phonon transition probabilities contain 
terms which are merely corrections to the one-phonon transition probability (\ref{Grule}). Thus, for transitions already 
triggered by a one-phonon process it may in some cases be reasonable to neglect, in a first approximation, these 
correction terms.

\subsection{Calculation of the transition probabilities}
\label{oneANDtwo}

The one-phonon transition probability \(\mathcal{R}^{1}(q^\prime,q)\) can easily be brought into the form of the golden 
rule~\cite{GKT80a},
\begin{align}
\mathcal{R}^{1}\left(q^{\prime},q \right)& =\frac{2\pi}{\hbar} \sum_{Q} 
G_{q,q^{\prime}}^{1} \left(Q\right)\left[G_{q,q^{\prime}}^{1}\left(Q\right)\right]^\ast \nonumber \\ 
&\times \left\{ n_{B}\left(\hbar \omega_{Q} \right) \delta \left(E_q-E_{q^{\prime}}+\hbar \omega_Q \right)  \right. 
\nonumber \\
&  \left. +  \left[ 1+ n_{B}\left(\hbar \omega_{Q} \right)\right]  
\delta \left(E_q-E_{q^{\prime}}-\hbar \omega_Q \right)_{}^{}  \right\} ,\label{1phononrateQ}
\end{align}
where the two terms in the curly brackets describe, respectively, the absorption and emission of a 
phonon.

To evaluate transition probabilities numerically we assume the phonon spectrum to be adequately represented 
by the Debye model. Sums over phonon wave numbers can thus be transformed into integrals according to 
\begin{align}
\sum_Q \dots = \frac{3 N_s}{\omega_D^3} \int_0^{\omega_D} \mathrm{d}\omega \omega^2 \dots \quad .
\label{Debyesumintegral}
\end{align}
Formula (\ref{onephononcompactrate}) in Appendix \ref{compactrates} gives the one-phonon transition probability in 
compact form as used in the numerical calculation.

%\subsection{Two-phonon rates}
%\label{twophononratessection}

The manipulation of the two-phonon transition probabilities is rather involved and cannot be reproduced entirely. 
In order to illustrate the necessary steps we take %\(\mathcal{R}^{10}\left(q^{\prime},q \right)\) the rate 
\begin{align}
\mathcal{R}^{10}\left(q^{\prime},q \right)&=\frac{2\pi}{\hbar} \sum_{s,s^\prime} \frac{e^{-\beta E_s}}{\sum_{s^{\prime\prime}} 
e^{-\beta E_{s^{\prime\prime}}}} \langle s^{\prime},q^{\prime}|V_{2} |s,q\rangle \nonumber \\
& \times \langle s,q|V_{1}^{\ast}G_{0}^{\ast}V_{1}^{\ast} |s^{\prime},q^{\prime}\rangle \nonumber \\ 
&\times  \delta (E_s -E_{s^\prime} +E_q -E_{q^\prime})  
\label{R10start}
\end{align} 
as a representative example. It contains both types of interactions: a simultaneous two-phonon interaction 
\(V_2\) and two successive one-phonon interactions \(V_1\) linked by a virtual intermediate state arising
from the iteration of the \(T\)-matrix. 

We begin the calculation with inserting into (\ref{R10start}) the expressions for \(V_1\), \(V_2\), and \( G_0\) 
as given by (\ref{vau1}), (\ref{vau2}), and (\ref{elphresolvent}), respectively. Inserting, furthermore,
the resolution of the identity over electron and phonon states,
%\(\mathcal{R}^{10}\) takes the form
\begin{widetext}
\begin{align}
\mathcal{R}^{10}\left(q^{\prime},q \right)=-\frac{2\pi}{\hbar} & \sum_{s,s^\prime} \frac{e^{-\beta E_s}}{\sum_{s^{\prime\prime}} e^{-\beta E_{s^{\prime\prime}}}}  \sum_{q_1,q_2} \sum_{s_1,s_2} \langle s^{\prime}| \sum_{Q_1,Q_2} G_{q^\prime,q}^{(2)}(Q_1,Q_2) \left( b_{Q_1}+b_{-Q_1}^\dagger\right)  \left( b_{Q_2}+b_{-Q_2}^\dagger\right) |s \rangle  \nonumber \\
& \times \langle s| \sum_{Q_3} \left[G_{q,q_1}^{(1)}(Q_3)\right]^\ast \left( b_{Q_3}^\dagger+b_{-Q_3} \right) |s_1\rangle \langle s_1,q_1| \frac{1}{E_s+E_q-H_0 -i\epsilon} |s_2,q_2\rangle  \nonumber \\
& \times\langle s_2 |\sum_{Q_4} \left[G_{q_2,q^\prime}^{(1)}(Q_4)\right]^\ast \left( b_{Q_4}^\dagger+b_{-Q_4} \right)  |s^{\prime}\rangle \delta (E_s -E_{s^\prime} +E_q -E_{q^\prime}) \text{ .}
\end{align}

\noindent
Using the two identities
\begin{align}
\delta(x)=\frac{1}{2\pi} \int_{-\infty}^\infty e^{ixt}\mathrm{d}t 
\qquad \text{and} \qquad
 \frac{1}{x-i\epsilon} = i\int_{-\infty}^0 e^{i(x-i\epsilon)\tau} \mathrm{d}\tau 
\end{align}
and the fact that the free resolvent is diagonal with respect to the electron-phonon states \(|q,s\rangle\) we obtain
\begin{align}
\mathcal{R}^{10}\left(q^{\prime},q \right)=&-\frac{2\pi}{\hbar} \sum_{q_1} \sum_{Q_1,Q_2,Q_3,Q_4} \frac{1}{2\pi} \int_{-\infty}^\infty \mathrm{d}t/\hbar \quad e^{i(E_q-E_{q^\prime}) \frac{t}{\hbar}} \quad  i \int_{-\infty}^0 \mathrm{d} \tau/\hbar \quad e^{i(E_q-E_{q_1}-i\epsilon)\frac{\tau}{\hbar}} G_{q^\prime,q}^{(2)} (Q_1,Q_2) \nonumber \\
&\times \left[G_{q_1,q}^{(1)}(Q_3)  G_{q^\prime,q_1}^{(1)} (Q_4) \right]^\ast  \frac{\sum_{s,s^\prime,s_1}  e^{-\beta E_s}} {\sum_{s^{\prime\prime}} e^{-\beta E_{s^{\prime\prime}}}} 
\langle s^\prime | \left( b_{Q_1}+b_{-Q_1}^\dagger\right)  \left( b_{Q_2}+b_{-Q_2}^\dagger\right) |s\rangle \nonumber \\
&\times \langle s| e^{i E_s \frac{t+\tau}{\hbar}} \left( b_{Q_3}^\dagger+b_{-Q_3} \right) e^{-i E_{s_1} \tau/\hbar} |s_1\rangle \langle s_1| \left( b_{Q_4}^\dagger+b_{-Q_4} \right) e^{-i E_{s^\prime} t/\hbar} |s^\prime \rangle \text{ ,}   
\label{R10expl}
\end{align}
%\end{widetext}
where all exponential factors containing electron energies have been placed in front of the phonon average. Employing 
\(\langle s| e^{iE_s t/\hbar}=\langle s| e^{i H^0_{ph} t/\hbar}\) and introducing \(v_Q=b_Q^{ }+b_{-Q}^\dagger\) the 
above expression becomes 
\begin{align}
\mathcal{R}^{10}(q^\prime,q)=&-\frac{2\pi}{\hbar} \sum_{q_1} \sum_{Q_1,Q_2,Q_3,Q_4} \frac{1}{2\pi} \int_{-\infty}^{\infty} \mathrm{d}t/\hbar \text{ } e^{i(\omega_q-\omega_{q^\prime})t} 
i\int_{-\infty}^0 \mathrm{d}\tau/\hbar\text{ } e^{i(\omega_q-\omega_{q_1}-i\epsilon) \tau} G_{q^\prime,q}^{(2)}(Q_1,Q_2)  \nonumber \\ 
& \times \left[G_{q^\prime,q_1}^{(1)}(Q_3) G_{q_1,q}^{(1)}(Q_4)\right]^\ast  \langle\langle v_{Q_3}^\dagger(t+\tau) v_{Q_4}^\dagger(t) v_{Q_1} v_{Q_2} \rangle\rangle \text{ ,}
\label{R10phononcorrelation}
\end{align}
where \( \langle \langle \dots \rangle\rangle= \sum_s e^{-\beta E_s}\langle s | \dots | s \rangle  
/ \sum_{s^{\prime \prime}} e^{-\beta E_{s^{\prime\prime}}} \) is the average over phonon states. The 
operator \(v_Q(t)\) evolves in time according to \(H^0_{ph}\). 
%the time dependence of the phonon creation 
%and annihilation operators is given by
%\begin{align}
%b_Q^\dagger (t)=e^{i\omega_Q t} b_Q^\dagger \quad \text{   and   } \quad b_Q (t)=e^{-i\omega_Q t} b_Q \text{ .} \label{freephononexplicittdep}
%\end{align}
Hence, the four-point phonon correlation function appearing in (\ref{R10phononcorrelation}) may be rewritten as 
\begin{align}
\langle \langle v_{Q_3}^\dagger (t+\tau) v_{Q_4}^\dagger (t) v_{Q_1} v_{Q_2}  \rangle \rangle  
&=e^{i\omega_{Q_3}(t+\tau)}e^{i\omega_{Q_4}t} \langle \langle b_{Q_3}^\dagger b_{Q_4}^\dagger   b_{Q_1}   b_{Q_2}  \rangle \rangle
+e^{i\omega_{Q_3}(t+\tau)}e^{-i\omega_{Q_4}t} \langle \langle b_{Q_3}^\dagger   b_{-Q_4}   b_{Q_1}   b_{-Q_2}^\dagger  \rangle \rangle\nonumber \\
&+e^{i\omega_{Q_3}(t+\tau)}e^{-i\omega_{Q_4}t} \langle \langle b_{Q_3}^\dagger   b_{-Q_4}  b_{-Q_1}^\dagger   b_{Q_2}  \rangle \rangle
+e^{-i\omega_{Q_3}(t+\tau)}e^{i\omega_{Q_4}t} \langle \langle b_{-Q_3}   b_{Q_4}^\dagger   b_{Q_1}  b_{-Q_2}^\dagger  \rangle \rangle \nonumber\\
&+e^{-i\omega_{Q_3}(t+\tau)}e^{i\omega_{Q_4}t} \langle \langle b_{-Q_3} b_{Q_4}^\dagger   b_{-Q_1}^\dagger   b_{Q_2} \rangle \rangle 
+e^{-i\omega_{Q_3}(t+\tau)}e^{-i\omega_{Q_4}t} \langle \langle b_{-Q_3} b_{-Q_4} b_{-Q_1}^\dagger   b_{-Q_2}^\dagger  \rangle \rangle~, 
\end{align} 
and further evaluated by forming all possible contractions. Using 
\begin{align}
\langle\langle b_{Q_1}^\dagger b_{Q_2} \rangle \rangle =\delta_{Q_1,Q_2} n_B(\hbar \omega_{Q_1}) 
\qquad \text{and} \qquad
\langle\langle b_{Q_1} b_{Q_2}^\dagger \rangle \rangle =\delta_{Q_1,Q_2} [1+n_B(\hbar \omega_{Q_1})]
\end{align}
and integrating over the times \(t\) and \(\tau\) finally yields 
%\begin{widetext}
\begin{align}
\mathcal{R}^{10}\left(q^{\prime},q \right)=&-\frac{2\pi}{\hbar} \sum_{q_1} \sum_{Q_1,Q_2}  G_{q^\prime ,q}^{2}\left(Q_{1},Q_1\right) \left[ G_{q_{1},q}^{1}\left(Q_{2}\right)    G_{q^\prime ,q_1}^{1}\left(Q_{2}\right) \right]^\ast \nonumber \\
& \times \left( 2   n_{B}\left(\hbar \omega_{Q_1} \right)  n_{B}\left(\hbar \omega_{Q_2} \right) \frac{\delta\left(E_{q}-E_{q^\prime}+\hbar\omega_{Q_{1}}+\hbar\omega_{Q_{2}} \right)}{E_q-E_{q_1}+\hbar\omega_{Q_1}-i\epsilon}   \right. \nonumber \\
& \left. \quad  +  \left[2 n_{B}\left(\hbar \omega_{Q_1} \right) n_{B}\left(\hbar \omega_{Q_2} \right)+ n_{B}\left(\hbar \omega_{Q_1} \right) \right] \frac{\delta\left(E_{q}-E_{q^\prime} \right)}{E_q-E_{q_1}+\hbar\omega_{Q_1}-i\epsilon}  \right. \nonumber  \\
& \left. \quad  +2 n_{B}\left(\hbar \omega_{Q_1} \right) \left[1+ n_{B}\left(\hbar \omega_{Q_2} \right)\right] \frac{\delta\left(E_{q}-E_{q^\prime}+\hbar\omega_{Q_{1}}-\hbar\omega_{Q_{2}} \right)}{E_q-E_{q_1}+\hbar\omega_{Q_1}-i\epsilon}\nonumber  \right.  
\\
& \left. \quad  +\left[2 n_{B}\left(\hbar \omega_{Q_1} \right) n_{B}\left(\hbar \omega_{Q_2} \right)+2 n_{B}\left(\hbar \omega_{Q_2} \right)+ n_{B}\left(\hbar \omega_{Q_1} \right)+1\right] \frac{\delta\left(E_{q}-E_{q^\prime} \right)}{E_q-E_{q_1}-\hbar\omega_{Q_1}-i\epsilon}  \right. \nonumber \\
& \left. \quad  + 2\left[1+ n_{B}\left(\hbar \omega_{Q_2} \right)\right] n_{B}\left(\hbar \omega_{Q_1} \right) \frac{\delta\left(E_{q}-E_{q^\prime}+\hbar\omega_{Q_{1}}-\hbar\omega_{Q_{2}} \right)}{E_q-E_{q_1}-\hbar\omega_{Q_2}-i\epsilon} \right.  \nonumber \\
& \left. \quad  +  2\left[1+ n_{B}\left(\hbar \omega_{Q_1} \right)\right] \left[1+ n_{B}\left(\hbar \omega_{Q_2} \right)\right] \frac{\delta\left(E_{q}-E_{q^\prime}-\hbar\omega_{Q_{1}}-\hbar\omega_{Q_{2}} \right)}{E_q-E_{q_1}-\hbar\omega_{Q_1}-i\epsilon} \right) \text{ .}\label{examplerate10b}
\end{align}
\end{widetext}

Similar expressions can be obtained for the other transition probabilities~\cite{Rafael09}. 
The formulae are all quite long. To gain more insight we classify two-phonon processes by the energy
difference they can bridge. As can be seen in the above example this is controlled by delta functions  
which, quite generally, appear in two-phonon transition probabilities with four different arguments:
\(\delta (E_q-E_{q^\prime})\),
\(\delta(E_q-E_{q^\prime}\pm\hbar \omega_Q)\),
\(\delta(E_q-E_{q^\prime}\pm (\hbar \omega_{Q_1}- \hbar \omega_{Q_2}))\), and
\(\delta (E_q-E_{q^\prime} \pm(\hbar \omega_{Q_1} + \hbar \omega_{Q_2}))\).
For the calculation of the transition probabilities we can drop all contributions proportional to 
\mbox{\(\delta (E_q-E_{q^\prime})\)} because in the one-dimensional model we are considering it implies no transition. 
%this means that \(q=q^\prime\), 
%but for transitions we have \(q\neq q^\prime\). 

In the Debye model, the maximum phonon energy is the Debye energy \(\hbar \omega_D\). Hence for terms in the 
transition probabilities  
that are proportional to \(\delta(E_q-E_{q^\prime}\pm\hbar \omega_Q) \) the maximal energy difference between the 
initial and final state of the electron cannot exceed one Debye energy. The two-phonon transition probabilities 
\(\mathcal{R}^3\), \(\mathcal{R}^5\), \(\mathcal{R}^6\), \(\mathcal{R}^7\), \(\mathcal{R}^{11}\), \(\mathcal{R}^{15}\), 
\(\mathcal{R}^{16}\) and \(\mathcal{R}^{17}\) have only contributions of this type. They are thus only 
corrections to the one-phonon transition probability \(\mathcal{R}^{1}\).  

Next, we consider terms proportional to \(\delta(E_q-E_{q^\prime}\pm (\hbar \omega_{Q_1}- \hbar \omega_{Q_2})) \). 
The energies of the two phonons \(\hbar \omega_{Q_1}\) and \(\hbar \omega_{Q_2}\) are both between zero and the 
Debye energy \(\hbar\omega_D\). As they appear with different signs in the delta function, the energy difference 
between the initial and final state of the electron can range from \(-\hbar\omega_D\) to \(\hbar\omega_D\). Thus 
these contributions do not allow to bridge levels that are farther apart than one Debye energy and are thus again
merely corrections to the one-phonon transition probability \(\mathcal{R}^{1}\). 

Finally, we look at the contributions to the transition probabilities proportional to 
\(\delta (E_q-E_{q^\prime} \pm(\hbar \omega_{Q_1} + \hbar \omega_{Q_2})) \). The energy difference that can be bridged 
by this type of process is between zero and two Debye energies. Up to an energy difference of one Debye energy these 
processes are again corrections to the one-phonon transition probability. But for energy differences between one and 
two Debye energies, these are the only processes that contribute to the transition probability. 
  
This analysis leads us to divide the two-phonon processes into two groups, one-Debye-energy transitions and 
two-Debye-energy transitions. One-Debye-energy transitions enable transitions between states that are at most 
one Debye energy apart, i.e. \(-1<\Delta_{q,q^\prime}<1\). Two-Debye-energy transitions enable transitions between 
states that are between one and two Debye energies apart, i.e. \(-2<\Delta_{q,q^\prime}<-1\) and 
\(1<\Delta_{q,q^\prime}<2\).

All the transition probabilities in the two phonon approximation contribute to one-Debye-energy transitions, but only 
the transition probabilities \(\mathcal{R}^9\), \(\mathcal{R}^{10}\), \(\mathcal{R}^{13}\) and \(\mathcal{R}^{14}\) 
contribute to two-Debye-energy transitions. Among the contributions to the one Debye energy transitions the golden rule 
transition probability is the only \textquotedblleft true\textquotedblright~one-phonon process. 

In the following, we assume that for one-Debye-energy transitions the one-phonon transition probability 
(\ref{1phononrateQ}) dominates the corrections from the two-phonon transition probabilities. Only when the one-phonon 
transition probability is zero, which is the case for 
two-Debye-energy transitions, we will take two-phonon processes into account. 
Hence, in our numerical calculation, we use for one-Debye-energy transitions the one-phonon transition 
probability,
\begin{align}
W_{q^\prime q}=\mathcal{R}^1(q^\prime,q)~,
\end{align}
and for two-Debye-energy transitions the two-phonon transition probability
\begin{align}
W_{q^\prime q}=\mathcal{\tilde{R}}^9(q^\prime,q)+2\mathrm{Re}\mathcal{\tilde{R}}^{10}(q^\prime,q)
+\mathcal{\tilde{R}}^{14}(q^\prime,q) \text{ ,}
\label{W2ph}
\end{align}
where the transition probabilities \(\mathcal{\tilde{R}}^9\), \(\mathcal{\tilde{R}}^{10}\) and \(\mathcal{\tilde{R}}^{14}\) 
denote those parts of \(\mathcal{R}^9\), \(\mathcal{R}^{10}\) and \(\mathcal{R}^{14}\) which 
give rise to two-Debye-energy transitions. They are given by equation (\ref{compactr9}), (\ref{compactr10}) and 
(\ref{compactr14}) in Appendix \ref{compactrates}. The transition probability \(\mathcal{\tilde{R}}^{13}\) does not appear 
explicitly. It is the complex conjugate to \(\mathcal{\tilde{R}}^{10}\) and thus subsumed in the 
second term on the rhs of Eq. (\ref{W2ph}).

\subsection{Regularization}
\label{Regularization}
The transition probabilities for the two-phonon processes contain, in the present form, divergences. Specifically within 
the two-Debye-energy approximation, which takes two-phonon processes into account only for transitions 
connecting (bound and unbound) surface states which are between one and two Debye energies apart, the 
transition probabilities \(\mathcal{\tilde{R}}^{10}\) and \(\mathcal{\tilde{R}}^{14}\) make trouble.

The divergences are artifacts of our one-dimensional model. They arise from the quantization of 
the electron motion perpendicular to the surface in conjunction with the harmonic approximation 
for the lattice. The former gives rise to arbitrarily sharp electronic energy levels for the bound 
states while the latter leads to infinite phonon lifetimes. Some divergent terms, for instance, 
\(I_{(2)}^3(2,1;2)\) and \(I_{(2)}^6\left(2,1;2,2\right)\) appearing, respectively, in the transition 
probabilities 
\(\mathcal{\tilde{R}}^{10}(2,1)\) and \(\mathcal{\tilde{R}}^{14}(2,1)\) (see Appendix \ref{compactrates}), 
can be traced back to the diagonal matrix element of the linear electron-phonon interaction (\ref{vau1}) 
and could thus be eliminated with a dressing transformation of the type used by Gortel and 
coworkers~\cite{GKT80b}. But other divergences, for instance, the one in the integral 
\(I_{(2)}^6(k^\prime,q;q_1,q_1)\) which appears in the rate \(\mathcal{\tilde{R}}^{14}(k^\prime,q)\) 
cannot be removed in that manner. We decided therefore to regularize the divergences of the transition 
probabilities by taking 
a finite phonon lifetime into account which works in all cases. The drawback of this procedure is that it 
turns divergences only into resonances, whose width is set by the phonon lifetime, which thus becomes an 
important additional material parameter.

In order to see how a finite phonon lifetime regularizes the transition probabilities, we recall bringing 
the transition probabilities into a numerically feasible form required to evaluate time integrals over products 
of time-dependent phonon two-point functions. For the transition probability \(\mathcal{\tilde{R}}^{10}\) we 
showed this explicitly (cf. Eq. (\ref{R10expl}) and the text which followed) but the same 
manipulations are necessary for the other transition probabilities~\cite{Rafael09}. Throughout we assumed that
the time evolution of the phonons is governed by the free phonon Hamiltonian \(H_{ph}^0\). As a result, 
the two-point functions acquired an undamped time dependence. In general, however, phonons interact, because of 
the anharmonicities in the lattice potential. A more realistic model would thus lead to phonon two-point functions 
whose time dependences are damped. Ultimately, the damping leads to divergence-free transition 
probabilities.

To account for the damping of phonons we imagine the retarded and advanced phonon Green functions to be 
given by 
\begin{align}
G^{R,A}(Q,\omega)=\frac{1}{\omega -\omega_Q\pm i\gamma_Q} \text{ ,}
\end{align}
where \(\gamma_Q\) is a decay constant arising from the phonon-phonon interaction and the upper (lower) 
sign corresponds to the retarded (advanced) Green function. Since the 
phonon four-point functions appearing in the two-phonon transition probabilities can be linked to these two  
functions, \(\gamma_Q\) can be incorporated into the expressions for the transition probabilities. Unfortunately, 
for the surfaces we are interested in little is known about the microphysics of phonons. We suggest
therefore to use a phenomenological estimate for \(\gamma_Q\) which utilizes material parameters which, 
at least in principle, could be measured~\cite{Klemens01}, 
\begin{align}
\gamma_Q = \frac{1}{\tau}=\frac{v}{l}=\frac{2 \gamma_G^2 \omega_Q^2 k_B T}{\mu V \omega_D} \text{ } \label{phenestgamma}
\end{align}
with \(\gamma_G\) the Gr\"uneisen parameter, \(V\) the volume per atom, and \(\mu\) the shear modulus.

In order to demonstrate how our regularization procedure works, we consider, again as an example, a four-point
function of the type appearing in (\ref{R10phononcorrelation}),
\begin{align}
\langle\langle & v_{Q_1}^\dagger(t_1) v_{Q_2}^\dagger(t_2) v_{Q_3}(t_3) v_{Q_4}(t_4) \rangle\rangle \nonumber \\
&= \langle\langle b_{Q_1}^\dagger(t_1) b_{Q_2}^\dagger(t_2) b_{Q_3}(t_3) b_{Q_4}(t_4)  \rangle\rangle \nonumber\\
&+\langle\langle b_{Q_1}^\dagger(t_1) b_{-Q_2}(t_2) b_{Q_3}(t_3) b_{-Q_4}^\dagger(t_4)  \rangle\rangle \nonumber\\ 
&+\langle\langle b_{Q_1}^\dagger(t_1) b_{-Q_2}(t_2) b_{-Q_3}^\dagger(t_3) b_{Q_4}(t_4)  \rangle\rangle \nonumber\\ 
&+\langle\langle b_{-Q_1}(t_1) b_{Q_2}^\dagger(t_2) b_{Q_3}(t_3) b_{-Q_4}^\dagger(t_4)  \rangle\rangle \nonumber\\ 
&+\langle\langle b_{-Q_1}(t_1) b_{Q_2}^\dagger(t_2) b_{-Q_3}^\dagger(t_3) b_{Q_4}(t_4)  \rangle\rangle \nonumber\\ 
&+\langle\langle b_{-Q_1}(t_1) b_{-Q_2}(t_2) b_{-Q_3}^\dagger(t_3) b_{-Q_4}^\dagger(t_4)  \rangle\rangle \text{ .} \label{typephonexvalue}
\end{align}
First, the four-point functions have to be broken up into two point functions. For the 
first term, e.g., this means
\begin{align}
\langle \langle & b_{Q_1}^\dagger(t_1)  b_{Q_2}^\dagger(t_2)  b_{Q_3}(t_3)  b_{Q_4}(t_4) \rangle \rangle \nonumber \\
&=\langle \langle b_{Q_1}^\dagger(t_1) b_{Q_3}(t_3) \rangle \rangle \langle \langle b_{Q_2}^\dagger(t_2) b_{Q_4}(t_4) \rangle \rangle \nonumber \\
&+\langle \langle b_{Q_1}^\dagger(t_1) b_{Q_4}(t_4) \rangle \rangle \langle \langle b_{Q_2}^\dagger(t_2) b_{Q_3}(t_3) \rangle \rangle \text{ .} \label{examplebreakup}
\end{align}
Because of translational invariance, the two-point functions are proportional to 
\(\delta_{Q_1,Q_2}\),  
%\begin{align}
%\langle \langle b_{Q_1}^\dagger(t_1) b_{Q_2}(t_2) \rangle \rangle = \delta_{Q_1,Q_2} \langle \langle b_{Q_1}^\dagger(t_1) b_{Q_2}(t_2) \rangle \rangle,
%\end{align}
even in the interacting case. 
The diagonal elements of the expectation values in (\ref{examplebreakup}) can be evaluated using the spectral theorem, 
%More specifically they are given by
\begin{align}
&\langle\langle b_Q^\dagger (t^\prime) b_Q(t) \rangle\rangle =\frac{1}{2\pi} \int_{-\infty}^{\infty} J_{b^\dagger b}(\omega) e^{-i\omega(t-t^\prime)} \mathrm{d}\omega
\label{bdaggerb}
\end{align}
and
\begin{align}
\langle\langle b_Q(t) b_Q^\dagger (t^\prime) \rangle\rangle =\frac{1}{2\pi} \int_{-\infty}^{\infty} J_{b^\dagger b}(\omega) e^{\beta \omega} e^{-i\omega(t-t^\prime)} \mathrm{d}\omega \text{  }
\label{bbdagger}
\end{align}
with the spectral function 
\begin{align}
J_{b^\dagger b}(\omega)=  \frac{i\left(G^R(\omega) - G^A(\omega)\right)}{e^{\beta \omega}-1} \text{ ,}
\end{align}
which contains the damping factor $\gamma_Q$ via the retarded and advanced phonon Green functions.  
Note, we neglect in the spectral function the contribution proportional to \(\delta(\omega)\). Since the 
formulae as they stand reproduce in the limit \(\gamma_Q \rightarrow 0\) the correct expressions for infinite 
phonon lifetime, we conclude that the neglect of the \(\delta(\omega)\) term is in this case justified. 
After integration, (\ref{bdaggerb}) and (\ref{bbdagger}) reduce, respectively, to 
\begin{align}
\langle\langle b_Q^\dagger (t^\prime) b_Q(t) \rangle\rangle =& n_B(\omega_Q) e^{-i\omega_Q(t-t^\prime)-\gamma_Q|t-t^\prime|}     
\end{align}
and
\begin{align}
\langle\langle b_Q(t) b_Q^\dagger (t^\prime) \rangle\rangle =& \left[1+n_B(\omega_Q)\right] e^{-i\omega_Q(t-t^\prime)-\gamma_Q|t-t^\prime|} \text{ .} 
\end{align}
Because of the damping factors the phonon four-point function in (\ref{R10phononcorrelation}) becomes
%which implies for the phonon four-point function in (\ref{R10phononcorrelation}) 
\begin{align}
\langle\langle & v_{Q_1}^\dagger(t+\tau_2)  v_{Q_2}^\dagger(t) v_{Q_3} v_{Q_4} \rangle\rangle \nonumber \\
&=2 n_B(\omega_{Q_1}) n_B(\omega_{Q_2}) e^{i\omega_{Q_1}(t+\tau_2)-\gamma_{Q_1} |t+\tau_2|} \nonumber \\
& \quad \times e^{i\omega_{Q_2}(t)-\gamma_{Q_2}|t|} \nonumber \\
& +2  \left[ 1+ n_B(\omega_{Q_1}) \right] \left[1+n_B(\omega_{Q_2}) \right] \nonumber \\ 
& \quad \times e^{-i \omega_{Q_1}(t+\tau_2)-\gamma_{Q_1}|t+\tau_2|} e^{-i\omega_{Q_2}t-\gamma_{Q_2}|t|} \nonumber \\
&+\dots \quad, \label{r10expphonlife} 
\end{align}
where the ellipsis stands for terms that do not allow for two-Debye-energy transitions. Performing finally in 
(\ref{R10phononcorrelation}) the integral over \(t\) and \(\tau\), with the phonon four-point function replaced by 
(\ref{r10expphonlife}), the dominant contribution, that is, the term giving rise to two-Debye-energy transitions
only, can be identified as 
\begin{align}
&\frac{1}{\pi}\frac{(\gamma_{Q_1}+\gamma_{Q_2})}{(\gamma_{Q_1}+\gamma_{Q_2})^2+(\omega_q-\omega_{q^\prime}\pm(\omega_{Q_1}+\omega_{Q_2}))^2} \nonumber \\ & \times \frac{\omega_q-\omega_{q_1}\pm\omega_{Q_1}}{(\omega_q-\omega_{q_1}\pm\omega_{Q_1})^2+\gamma_{Q_1}^2} \text{ ,}
\label{interestterm}
\end{align}
and, for small decay constants \(\gamma_Q\), approximated by
\begin{align}
&\delta(\omega_q-\omega_{q^\prime}\pm(\omega_{Q_1}+\omega_{Q_2})) \nonumber \\ 
&\times \frac{\omega_{q^\prime}-\omega_{q_1}\mp\omega_{Q_2}}{(\omega_{q^\prime}-\omega_{q_1}\mp\omega_{Q_2})^2+(2\gamma_{Q_1}+\gamma_{Q_2})^2} \text{ .}
\end{align}
Putting finally everything together, the corrected, divergence-free transition probability \(\mathcal{\tilde{R}}^{10}\) becomes 
\begin{widetext}
\begin{align}
\mathcal{\tilde{R}}^{10}(q^\prime,q)&=-\frac{2\pi}{\hbar} \sum_{q_1} \sum_{Q_1,Q_2} G_{q^\prime,q}^{(2)}(Q_1,Q_1) \left[G_{q_1,q}^{(1)}(Q_2)G_{q^\prime,q_1}^{(1)}(Q_2)\right]^\ast  \nonumber\\
& \times \left( 2 n_B(\hbar\omega_{Q_1}) n_B(\hbar\omega_{Q_2}) \delta(E_q-E_{q^\prime}+\hbar\omega_{Q_1}+\hbar\omega_{Q_2})  g(E_{q^\prime}-E_{q_1}-\hbar\omega_{Q_2},2\hbar\gamma_{Q_1}+\hbar\gamma_{Q_2}) \right.\nonumber \\[0.1cm]
& \left. + 2 [1+n_B(\hbar\omega_{Q_1})][1+n_B(\hbar\omega_{Q_2})] \delta(E_q-E_{q^\prime}-\hbar\omega_{Q_1}-\hbar\omega_{Q_2})  g(E_{q^\prime}-E_{q_1}+\hbar\omega_{Q_2},2\hbar\gamma_{Q_1}+\hbar\gamma_{Q_2}) \right) \text{ ,}
\label{R10tildeReg}
\end{align}
where we have used the abbreviation
\begin{align}
g(a,\gamma_a)=\frac{a}{a^2+\gamma_a^2} \text{ .}
\end{align}

A similar analysis can be performed for the transition probability \(\mathcal{\tilde{R}}^{14}\). 
Introducing the function
\begin{align}
f(a,b,\gamma_a,\gamma_b)=\frac{ab+\gamma_a \gamma_b}{(a^2+\gamma_a^2)(b^2+\gamma_b^2)} \text{ ,} 
\end{align}
the corrected, divergence-free transition probability \(\mathcal{\tilde{R}}^{14}\) is then given by
\begin{align}
\mathcal{\tilde{R}}^{14}=&\frac{2\pi}{\hbar} \sum_{q_1,q_2} \sum_{Q_1,Q_2} G_{q,q_1}^{(1)}(Q_1) G_{q_1,q^\prime}^{(1)}(Q_1) \left[G_{q,q_2}^{(1)}(Q_2) G_{q_2,q^\prime}^{(1)}(Q_2)\right]^\ast \nonumber\\
& \times \left( n_B(\hbar\omega_{Q_1}) n_B(\hbar\omega_{Q_2}) \delta(E_q-E_{q^\prime}+\hbar\omega_{Q_1}+\hbar\omega_{Q_2}) \right. \nonumber \\
 & \left. \quad \times f(E_{q_2}-E_{q^\prime}+\hbar\omega_{Q_2},E_{q_1}-E_{q^\prime}+\hbar\omega_{Q_1},2\hbar\gamma_{Q_1}+\hbar\gamma_{Q_2},\hbar\gamma_{Q_1}+2\hbar\gamma_{Q_2}) \right. \nonumber\\
 & \left. + n_B(\hbar\omega_{Q_1}) n_B(\hbar\omega_{Q_2}) \delta(E_q-E_{q^\prime}+\hbar\omega_{Q_1}+\hbar\omega_{Q_2}) \right.  \nonumber\\
 & \left.  \quad \times  f(E_{q_2}-E_{q^\prime}+\hbar\omega_{Q_2},E_{q_1}-E_{q^\prime}+\hbar\omega_{Q_2},2\hbar\gamma_{Q_1}+\hbar\gamma_{Q_2},2\hbar\gamma_{Q_1}+\hbar\gamma_{Q_2}) \right. \nonumber\\
 & \left. +[1+ n_B(\hbar\omega_{Q_1})][1+ n_B(\hbar\omega_{Q_2})] \delta(E_q-E_{q^\prime}-\hbar\omega_{Q_1}-\hbar\omega_{Q_2}) \right.\nonumber \\
 & \left. \quad \times f(E_{q_2}-E_{q^\prime}-\hbar\omega_{Q_2},E_{q_1}-E_{q^\prime}-\hbar\omega_{Q_1},2\hbar\gamma_{Q_1}+\hbar\gamma_{Q_2},\hbar\gamma_{Q_1}+2\hbar\gamma_{Q_2}) \right.\nonumber\\
 & \left. + [1+ n_B(\hbar\omega_{Q_1})][1+ n_B(\hbar\omega_{Q_2})] \delta(E_q-E_{q^\prime}-\hbar\omega_{Q_1}-\hbar\omega_{Q_2}) \right.\nonumber \\
 & \left.  \quad \times  f(E_{q_2}-E_{q^\prime}-\hbar\omega_{Q_2},E_{q_1}-E_{q^\prime}-\hbar\omega_{Q_2},2\hbar\gamma_{Q_1}+\hbar\gamma_{Q_2},2\hbar\gamma_{Q_1}+\hbar\gamma_{Q_2}) \right) \text{ .}
\label{R14tildeReg}
\end{align}
\end{widetext}

The transition probability \(\mathcal{\tilde{R}}^{9}\) has no divergence and hence requires no regularization.
Equations(\ref{r10compactcorr}) and (\ref{r14compactcorr}) in Appendix \ref{compactrates} give the final 
expressions for the corrected, divergence-free two-phonon transition probabilities as used in the numerical calculation.

\section{Results}
\label{Results}
\begin{table}[t]
\caption{Material parameters for graphite. 
%To obtain numerical results for 
%$\tau_{\rm e}^{-1}$ we used this set of parameters except for Debye temperature which we varied 
%to simulate surface potentials of different depths.
}
\center
\begin{tabular}{l l l l l}
\hline
Debye temperature & \(\quad\) & \(T_D\) & \(\quad\) & \(2500 K\) \\
Dielectric constant & \(\quad\) & \(\epsilon_s\) & \(\quad\) & \(13.5\) \\
TO phonon energy & \(\quad\) & \(\hbar \omega_T\) & \(\quad\) & \(170 \text{meV}\)  \\
Gr\"uneisen parameter & \(\quad\) &  \(\gamma_G\) & \(\quad\) & \(1.7\) \\
Shear modulus & \(\quad\) & \(\mu  \) & \(\quad\) & \(5\) GPa  \\
\hline
\end{tabular}
\label{materialtable}
\end{table}

We now insert the one- and (regularized) two-phonon transition probabilities into the rate equation (\ref{desrateeq}) 
to investigate phonon-induced desorption for an electron image-bound to a dielectric surface. The 
two-phonon approximation is of course only applicable to dielectrics which have a two-phonon deep potential,
for instance, graphite (see discussion in Section \ref{Introduction}). For graphite, the material parameters
required for a calculation of the desorption time are summarized in table \ref{materialtable}. Apart from 
the Debye temperature, which we varied to study the dependence of the desorption time on the potential depth, 
all numerical results were obtained for these parameters.

\subsection{One-phonon transition probability}

To set the stage, we start with presenting results for the desorption time \(\tau_{\rm e}\) calculated with 
one-phonon processes only, although these times do not apply to any of the dielectric materials we mentioned.

\begin{figure}[t]
\includegraphics[width=\linewidth]{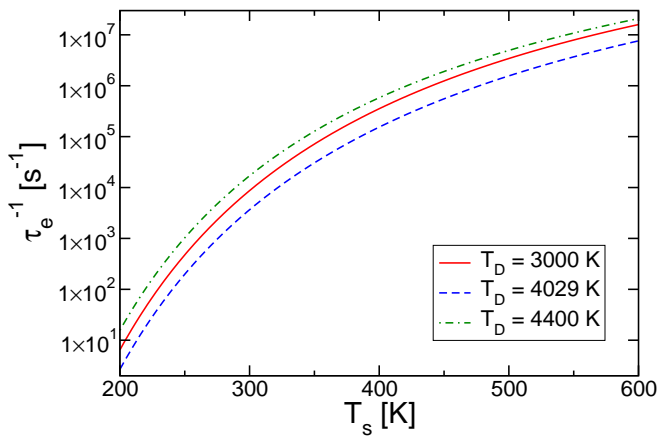}
\caption{Inverse desorption time \(\tau_{e}^{-1}\) in the one-phonon approximation 
as a function of the surface temperature \(T_S\) for a one-phonon deep 
potential (\(T_D=3000K\), \(T_D=4029K\)) and a shallow potential (\(T_D=4400K\)).}
\label{1phsurf}
\end{figure}

First, we discuss the dependence of the desorption time on the surface temperature. If 
\(\tau_{\rm e}\) is much larger than the time the electron needs to thermally equilibrate with the surface, 
prior to desorption, electron surface states are populated according to $n_q\sim\exp[-E_q/kT_s]$. Since 
for $q\ge 2$, $-E_q/kT_s\ll -E_1/kT_s$ this means that the electron basically desorbs from the lowest surface 
bound state. 
Desorption requires therefore transitions from the lowest bound state to the upper bound states and finally to 
the continuum. They entail the absorption of a phonon. The likelihood of which is, according to (\ref{1phononrateQ}),
proportional to the Bose-distribution \(n_B\) and thus increases strongly with temperature. This is reflected in 
Fig.~\ref{1phsurf} showing that an increase of the surface temperature leads to an increase 
of \(\tau_{\rm e}^{-1}\) over several orders of magnitude. Variations of the Debye temperature in contrast do not 
change the strong temperature dependence significantly.

We now move on to study the effects of the potential depth on desorption. In 
Section \ref{Desorption from many bound states} we explained how the 
potential depth can be classified by the maximum phonon energy, the Debye energy \(\hbar\omega_D\). The 
relative depth of the potential can be changed easily by tuning the Debye energy whilst keeping the absolute 
potential depth constant. This is advantageous from the technical perspective since the cumbersome calculation 
of the electronic matrix elements \(Z_{qq^\prime}^n\) does not have to be repeated. In order to keep the level 
of phonon excitation constant while the Debye temperature is varied we set the inverse surface temperature
constant,
\begin{align}
\delta=\frac{\hbar\omega_D}{k_BT_S}~.
\end{align}

\begin{figure}[t]
\includegraphics[width=\linewidth]{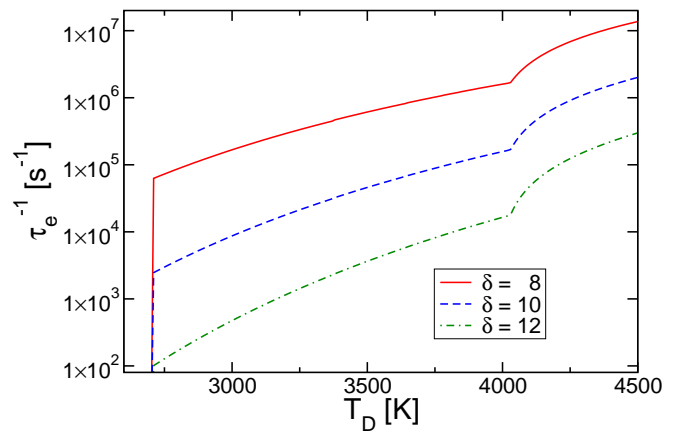}
\caption{Inverse desorption time \(\tau_{e}^{-1}\) in the one-phonon approximation as a function of the Debye 
temperature \(T_D\) for different inverse temperatures \(\delta\). Tuning \(T_D\) the potential can be 
made shallow (\(T_D > 4029 K \)), one-phonon deep (\(2707 K < T_D <4029 K \)) or two-phonon deep 
(\(T_D<2707 K \)) for which desorption by one-phonon processes is impossible. }
\label{1phtdebye}
\end{figure}

Figure~\ref{1phtdebye} shows $\tau_{\rm e}^{-1}$ depending on the Debye temperature $T_D$. The lower $T_D$, 
the larger the effective potential depth. For a one-phonon deep potential, $T_D>2707~K$, the lowest level 
is coupled to at least one other bound state by a one-phonon process. In this region the calculation of 
$\tau_{\rm e}^{-1}$ using one-phonon processes only is applicable and leads to an increase of 
$\tau_{\rm e}^{-1}$ with increasing $T_D$, that is, with decreasing effective potential depth. This 
is what one would expect as it should be easier to get the energy required
to bridge a smaller energy difference than a larger one. For a shallow potential, $T_D>4029~K$, the lowest 
bound state can be emptied directly to the continuum. This leads to a substantial increase of 
$\tau_{\rm e}^{-1}$ signaled by the kink. The main conclusion at this point is that direct transitions which 
are only possible for a shallow potential are more effective than a cascade of transitions which are the only 
means of emptying a deep potential.

\begin{figure}[t]
\includegraphics[width=\linewidth]{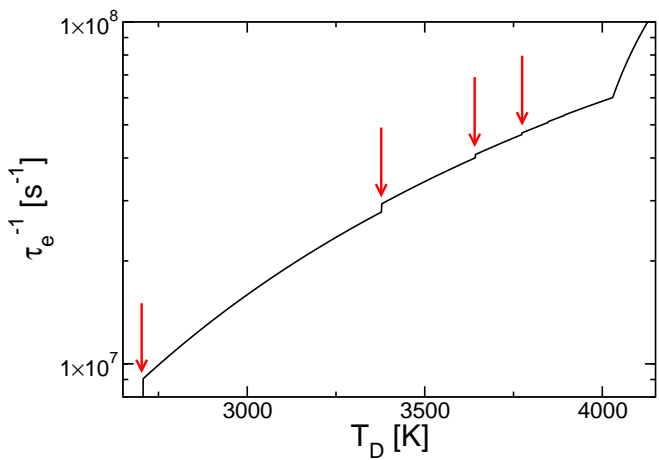}
\caption{Inverse desorption time \(\tau_e^{-1}\) in the one-phonon approximation as a function of the Debye 
temperature for a high 
surface temperature (\(\delta = 5\)). The small steps under the red arrows coincide with the onset of 
transitions from the lowest to the second, third, fourth, fifth, etc. bound state.}
\label{1phtdebyebis}
\end{figure}

To gain further insight, we identify the most relevant cascade. The one-phonon deep potential, for which the 
transition between the lowest two bound states is a one-phonon process, can be subdivided further, depending on the 
accessibility of the higher bound states. For the one-phonon deep potential the lowest level does not couple to the 
continuum directly, but it is coupled to at least the second bound state. Transitions between the lowest bound state 
and the third, fourth, fifth, \dots bound state may or may not be possible. For example,
for \(\Delta_{1,2} > -1 > \Delta_{1,3}\) only the second bound state,
for \(\Delta_{1,3} > -1 > \Delta_{1,4}\) the second and third bound state, and hence
for \(\Delta_{1,n} > -1 > \Delta_{1,n+1}\) the second to \(n^\text{th}\) bound state can be reached from the lowest 
state.
These accessibility thresholds mark the opening of new desorption channels when the potential depth is reduced.  
Figure~\ref{1phtdebyebis} shows that when the second, third, or fourth level becomes available from the lowest level, 
\(\tau_{\rm e}^{-1}\) increases suddenly, although these steps are small. We deduce that the first leg of the cascade 
to the continuum is predominantly the transition to the second level. 

\begin{figure}[t]
\includegraphics[width=\linewidth]{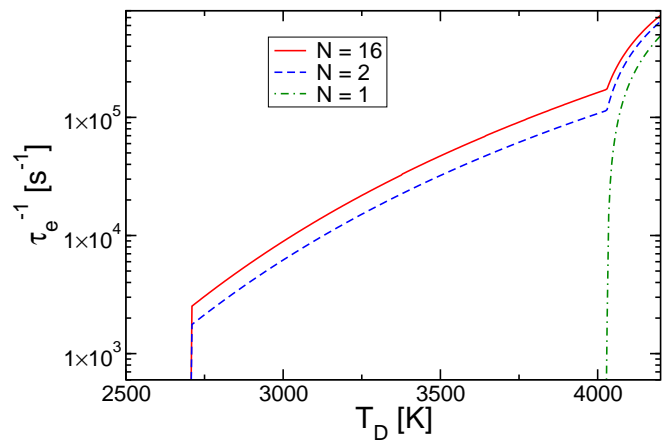}
\caption{Inverse desorption time \(\tau_e^{-1}\) in the one-phonon approximation as a function of the Debye 
temperature for \(\delta=10\) 
calculated with different numbers of bound states \(N\). For \(N=1\) desorption occurs only due to direct 
transition to the continuum which dominates the rate for a shallow potential (\(T_D > 4029~K\)). For \(N>1\)
cascade transitions allow for desorption from a one-phonon deep potential (\(2707~K < T_D <4029~K \)). In this 
case the second bound state gives the most important contribution. }
\label{1phN}
\end{figure}

The question arises as to how important the higher bound states \(n=3,4,5\dots\) are as intermediate steps for the 
second leg of a desorption cascade. To investigate this we calculated \(\tau_{\rm e}^{-1}\) with different numbers of 
bound states. 
The image potential has an infinite sequence of bound states, yet for practical reasons we can take only a finite number 
of bound states into account. Figure~\ref{1phN} confirms that adding higher bound states to the calculation 
\(\tau_{\rm e}^{-1}\) saturates quickly. Neglecting all but a few bound states, say seven, is therefore justified. 
With two bound states the value of \(\tau_{\rm e}^{-1}\) amounts already to about two thirds of the value obtained with 
sixteen bound states. Hence the second leg 
of the dominant desorption channel is a direct transition from the second bound state into the continuum. The reason for 
the importance of the second level lies in the matrix element \(Z^2_{q,q^\prime}\) which is large for low bound states for 
which more probability density is concentrated near the surface where the dynamic perturbation inducing desorption is 
strongest.

\subsection{Two-phonon transition probability}

Under the assumption that the true one-phonon transition probability  
(\ref{1phononrateQ}) dominates for one-Debye-energy transitions the corrections coming from the two-phonon 
transition probabilities, the latter need only be considered for two-Debye-energy transitions, for which 
the transition probabilities would be zero otherwise. All the data presented in this subsection were obtained 
within this approximation. 

For $T_D=2500 K$, the numerical results apply to an electron image-bound to graphite, which 
has, in our notation, a two-phonon deep surface potential. Indeed, using the dynamically corrected image 
potential, we find for the lowest two image states of graphite \(E_1=-0.347~eV \) and \(E_2= -0.114~eV\). 
Hence, \(\Delta_{12}=-1.06\) implying \(-1>\Delta_{12}>-2\) and thus a two-phonon deep surface potential. 
Within our main assumption that the transition probability corresponding to the minimum number of phonons 
needed to open for the first time a particular transition is the dominant one, the 
two-phonon approximation is sufficient for graphite; $n$-phonon processes with $n\ge 3$ should yield 
only small corrections. 

%As pointed out in the introduction, image states on a graphite surface have been experimentally 
%detected~\cite{LMT99} but the binding energy, \(E_1^{\rm exp}\approx -0.85~eV\), is closer to the value of 
%the classical image potential than to the one obtained from the more realistic dynamically corrected image 
%potential (see previous paragraph). Further investigations, ideally including studies of trapping and 
%de-trapping of external electrons, seem to be thus clearly necessary.

\begin{figure}[t]
\includegraphics[width=\linewidth]{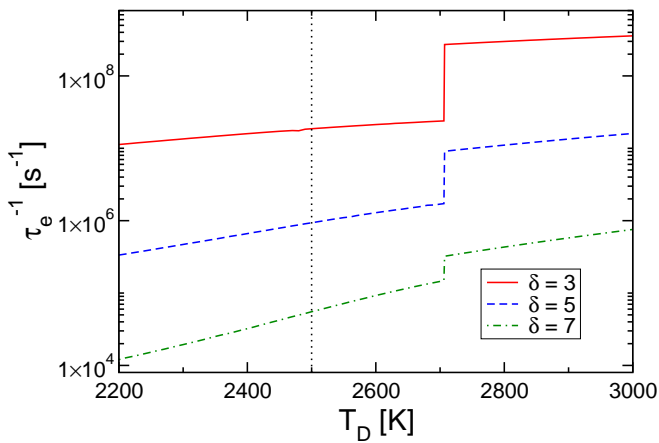}
\caption{Inverse desorption time \(\tau_{e}^{-1}\) in the two-phonon approximation as a function of the Debye 
temperature \(T_D\) for a two-phonon 
(\(T_D<2707~K\)) and a one-phonon deep potential (\(T_D>2707~K\)) for different surface 
temperatures \(\delta\). At \( T_D=2707~K \), the onset of one-phonon transitions between the lowest 
two states, \(\tau_{\rm e}^{-1}\) increases considerably. Data for $T_D=2500~K$ apply to graphite 
(thin vertical line).}
\label{2phtdebye}
\end{figure}

Figure~\ref{2phtdebye} concerns once more the dependence of \(\tau_{\rm e}^{-1}\) on the Debye temperature, 
but this time also for Debye temperatures leading to two-phonon deep potentials ($T_D<2707~K$). Using 
one-phonon transition probabilities only \(\tau_{\rm e}^{-1}\) would drop from a finite value to zero when 
the one-phonon deep potential (\(\Delta_{12} > -1 \)) becomes two-phonon deep (\(\Delta_{12}<-1\)). This 
happens at $T_D=2707~K$. Including two-phonon transition
probabilities leads to a finite \(\tau_{\rm e}^{-1}\) even for two-phonon deep potentials. The data 
in  Fig.~\ref{2phtdebye} for $T_D=2500~K$ apply to graphite (thin vertical line). For instance, for 
$\delta=7$, that is, \(T_s\approx 360 K\), we find \(\tau_{\rm e}^{-1} \approx 5\cdot 10^4~s^{-1}\) 
and hence a desorption time \(\tau_{\rm e}\approx 2\cdot 10^{-5}~s\).

If a two-phonon deep potential is made shallower so that it becomes one-phonon deep at \(\Delta_{12}=-1\), 
the stronger one-phonon transitions set in and \(\tau_{\rm e}^{-1}\) increases considerably. For high surface 
temperatures, for instance for \(\delta=5\), \(\tau_{\rm e}^{-1}\) increases about fivefold. Hence, for high 
surface temperatures the one-phonon transition probabilities dominate their two-phonon corrections as expected. 
For lower surface temperatures, however, the increase in \(\tau_{\rm e}^{-1}\) at the onset of one-phonon 
transitions becomes smaller. For instance, for \(\delta=7\) it amounts only to a factor of two. For very low surface 
temperatures, \(\tau_{\rm e}^{-1}\) even drops at the threshold, e.g. for \(\delta=20\) by about 40\% 
(not shown in Fig. \ref{2phtdebye}). In this case our assumption that the one-phonon transition probability dominates 
its two-phonon corrections is clearly not justified. An accurate calculation of \(\tau_{\rm e}^{-1}\) for a one-phonon 
deep potential near \(\Delta_{12}=-1\) requires therefore a calculation with all two-phonon transition 
probabilities included.

Within our model for the polarization-induced surface potential, graphite is very close to the \(\Delta_{12}=-1\)
threshold. The neglected two-phonon corrections to the one-phonon transition probabilities, however, would 
be only critical if \(\Delta_{12}\) were slightly larger than \(-1\), not smaller, as it is in fact the 
case. Despite the approximations, we expect our numerical results to be reasonable  
for graphite, especially at higher temperatures, where the resonances of the regularized two-phonon transition
probabilities, which are the reminiscences of the divergences of the original transition probabilities, are washed 
out making the transition probabilities rather robust against small changes in model parameters.

\begin{figure}[t]
\includegraphics[width=\linewidth]{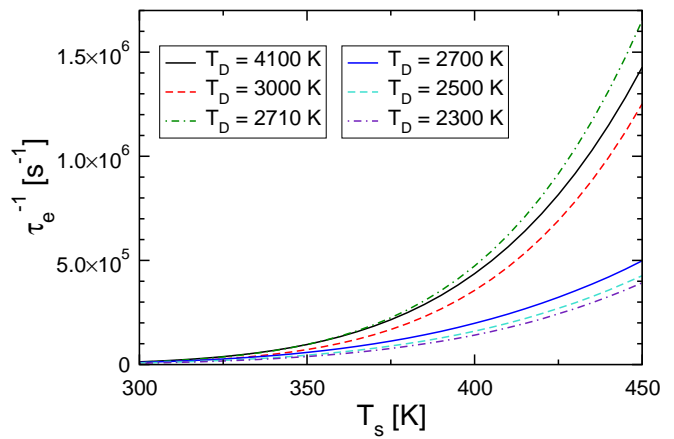}
\caption{Inverse desorption time \(\tau_{e}^{-1}\) in the two-phonon approximation as a function of the surface temperature 
for different potential depths. For high surface temperatures desorption from a two-phonon deep potential 
(\(T_D=2700~K,2500~K,2300~K\)) is significantly slower than desorption from a one-phonon deep potential 
(\(T_D=2710~K,3000~K\)) or a shallow potential (\(T_D=4100~K\)).}
\label{2phsurf}
\end{figure}

Figure~\ref{2phsurf} compares the dependence of \(\tau_{\rm e}^{-1}\) on the surface temperature for 
the potential depths shallow, one-phonon deep, and two-phonon deep, as realized by different values
for the Debye temperature. For all potential depths \(\tau_{\rm e}^{-1}\) increases 
with surface temperature. The increase of \(\tau_{\rm e}^{-1}\) for shallow and one-phonon 
deep potentials is about the same and significantly steeper than for two-phonon 
deep potentials. For high temperatures, therefore, desorption from two-phonon deep potentials lags 
behind desorption from one-phonon deep potentials.

\begin{figure}[t]
\includegraphics[width=\linewidth]{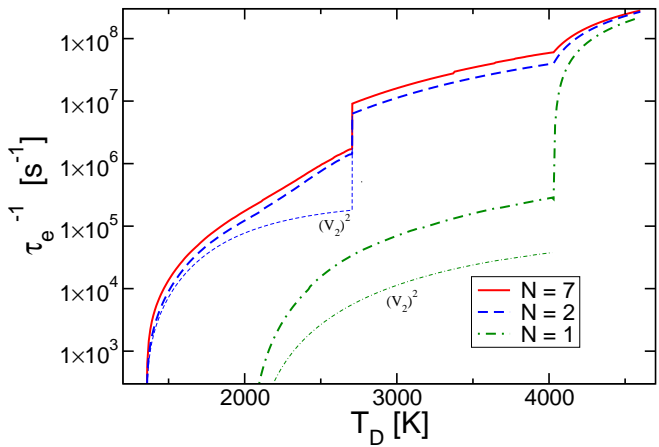}
\caption{Inverse desorption time \(\tau_{e}^{-1}\) in the two-phonon approximation for $\delta=5$ as a function of the Debye temperature \(T_D\) calculated with different numbers of bound states \(N\). Above \(T_D=2010~K\) the continuum is accessible from the lowest bound state by two-phonon processes, above  \(T_D=4029~K\) by one-phonon processes. For \(T_D < 2707~K\) the potential is two-phonon deep, for \(2707~K<T_D <4029~K\) it is one-phonon deep and above \(T_D=4029~K\) it is shallow. For the thin lines labeled with \((V_2)^2\) the two-phonon process has been calculated using \(\mathcal{R}^9\) only. }
\label{2phN}
\end{figure}

For the calculation with one-phonon transitions included only we identified the second bound state as the most important 
intermediate state for the desorption cascade. We now study the role of intermediate bound states when two-phonon processes 
are taken into account. The goal is again to reveal the relative importance of direct desorption vs. desorption via 
cascades. Figure~\ref{2phN} shows that \(\tau_{\rm e}^{-1}\) saturates quickly with the number of bound states considered. 
Calculating \(\tau_{\rm e}^{-1}\) with only the two lowest bound states included gives essentially the correct result. 
In the case
of the one-phonon calculation, we inferred from the fact that we need only two bound states, that the transition from the 
first to the second state is the most important one. Within the two-phonon calculation, however, the interpretation 
is not that simple because an additional bound state besides the lowest one has two influences: Firstly, it makes  
cascade transitions with an intermediate bound state possible, secondly it alters the transition probability for direct 
transitions from the lowest bound state to the continuum because it also acts as a virtual intermediate state in the 
contributions to the transition probabilities that stem from the iteration of the \(T\)-matrix. 
Although the direct two-phonon transition probability from the lowest bound state to the continuum is modified it does 
not increase significantly with additional virtual intermediate bound states. Hence, the cascade transitions are by far 
more important than the modified direct transitions and make up almost the whole transition probability on their own.

In addition to the identification of the most efficient desorption channel, Fig.~\ref{2phN} enables us to compare our 
results with the ones obtained by Gortel, Kreuzer and Teshima~\cite{GKT80b}. Their Figs.~1--3, 5 and 6 show 
\(\tau_{\rm e}^{-1}\) for a single bound state as a function of the bound state energy, whereas our Fig.~\ref{2phN} 
shows \(\tau_{\rm e}^{-1}\) as a function of the Debye temperature, which is proportional to the inverse of the 
potential depth. Hence, apart 
from scaling, Fig.~\ref{2phN} is mirror inverted compared to their figures. For \(N=1\) our approach applies to 
desorption from a single bound state, the situation studied by Gortel and coworkers. Despite the differences in the 
surface potential, arising from the fact that we are concerned with physisorption of an electron and Gortel et al. 
with physisorption of an atom, we also find that for potentials with depths allowing one-phonon 
transitions to the continuum desorption is much faster than 
for potentials whose depths require a two-phonon process for the transition from the bound state to the continuum. This 
results in a steep drop of \(\tau_{\rm e}^{-1}\) at \(T_D<2707~K \), when the one-phonon deep potential becomes 
two-phonon deep. In contrast 
to  Gortel et al., however, we can include in the two-phonon calculation the other bound states. Then, for deep 
potentials, the stronger cascades set in and lead to a substantial increase of \(\tau_{\rm e}^{-1}\).

Lastly we look at the relative importance of the two-phonon processes arising, respectively, from the expansion 
of the perturbation (\ref{tmatiteration}) and the iteration of the \(T\)-matrix (\ref{uexpanpotential}). A 
two-phonon process can be simultaneous, as encoded in \(V_2\), or successive, as 
described by \(V_1 G_0 V_1\). Hence, the total two-phonon 
transition probability (\ref{W2ph}) contains a contribution without virtual intermediate states, 
symbolically denoted by \((V_2)^2\) (see (\ref{rr9})) and two contributions with virtual 
intermediate states, symbolically denoted by \((V_1)^2V_2\) and \((V_1)^4\) (see (\ref{rr10}), (\ref{rr13}), 
and (\ref{rr14})). 

The inverse desorption time obtained from a calculation where only two-phonon transitions due to 
\( \mathcal{\tilde{R}}^{9}\), that is, due to the process \((V_2)^2\) have been included 
is shown by the thin green and blue lines 
in Fig.~\ref{2phN}. For the direct transition from the lowest bound state (\(N=1\), thin green line) to the 
continuum the process \((V_2)^2\) is always dominated by the processes \((V_1)^2V_2\) and \((V_1)^4\), as can 
be deduced by comparing the thick and thin green lines. For the more important cascade (\(N=2\), thin
blue line), however, the situation is more subtle. The processes \((V_1)^4\) and \((V_1)^2V_2\),
exhibiting resonances at \(\Delta_{12}=-1\) (recall Eqs. (\ref{R10tildeReg}) and (\ref{R14tildeReg}) for the 
regularized two-phonon transition probabilities \(\mathcal{\tilde{R}}^{10}\) and 
\(\mathcal{\tilde{R}}^{14}\), respectively) are important only near 
\(\Delta_{12}=-1\), that is, in the vicinity of \(T_D=2707~K\). 
Far away from \(T_D=2707~K\), it is in fact the process \((V_2)^2\) which gives the 
main contribution, as can be seen from the thin and thick blue lines in Fig.~\ref{2phN}. 

We conclude therefore that depending on the type of the desorption process (direct vs. indirect via cascades), 
the energy difference between initial and final state, and the energy of the virtual intermediate states either 
of the two-phonon processes, \((V_2)^2\), \((V_1)^2V_2\) and \((V_1)^4\), may be the most important one and 
neither can thus be neglected.

\section{Conclusions}
\label{Conclusions}

We investigated phonon-mediated desorption of an image-bound electron from dielectric surfaces using a 
quantum-kinetic rate equation for the occupancies of the bound surface states for the electron. To avoid 
the unphysical divergence of the classical image potential, we included the recoil experienced by the 
electron when it couples to the dipole-active modes responsible for the polarization-induced interaction
between the electron and the surface. Due to the coupling to bulk acoustic phonons an electron
initially occupying bound surface states may desorb when it gains enough energy to either directly
reach an extended state or to successively climb up the ladder of bound states. To allow for an 
efficient calculation of the electronic matrix elements entering the transition probabilities in the 
quantum-kinetic rate equation, we derived asymptotic approximations for the electron wave functions 
and matrix elements.     

For the dielectric materials relevant for bounded gas discharges (graphite, silicon oxide, aluminum oxide)
or electron emitting devices (Cs-doped glass and GaAs heterostructures) the energy spacing of at
least the two lowest image states is larger than the Debye energy. Hence, provided the surface temperatures 
are low enough for the electron to basically desorb from the lowest bound state, phonon-induced desorption
has to occur for these materials via multi-phonon processes, as they arise from the expansion of the electron-surface 
interaction potential with respect to the displacement field (originating from acoustic phonons) and the iteration 
of the \(T\)-matrix encoding the successive scattering of the external electron on the displacement field. 
Desorption channels involving internal electronic degrees of freedom are closed because the 
typical surface temperatures are too low for exciting these high-energy modes.

We presented results for a two-phonon deep surface potential, where the energy difference between
the lowest two bound states is between one and two Debye energies. Classifying two-phonon processes by the energy
difference they allow to bridge in one- and two-Debye-energy transitions, we included two-phonon transition 
probabilities only for two-Debye-energy transitions, that is, for transitions where the one-phonon transition
probability vanishes. We regularized moreover spurious singularities in the two-phonon transition probabilities
by taking a finite phonon lifetime into account.
\begin{figure}[t]
\includegraphics[width=\linewidth]{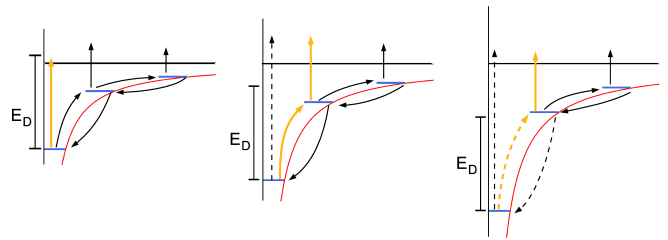}
%\begin{minipage}[b]{0.32\linewidth}
%\includegraphics[width=\linewidth]{a.eps}
%\vspace{0.8cm}
%\end{minipage}
%\begin{minipage}[b]{0.32\linewidth}
%\includegraphics[width=\linewidth]{b.eps}
%\vspace{0.5cm}
%\end{minipage}
%\begin{minipage}[b]{0.32\linewidth}
%\includegraphics[width=\linewidth]{c.eps}
%\end{minipage}
%\begin{minipage}{0.32\linewidth}
%\includegraphics[width=\linewidth]{2shuffle1.eps}
%\end{minipage}
%\begin{minipage}{0.32\linewidth}
%\includegraphics[width=\linewidth]{2shuffle2.eps}
%\end{minipage}
%\begin{minipage}{0.32\linewidth}
%\includegraphics[width=\linewidth]{2shuffle3.eps}
%\end{minipage}
\caption{Desorption channels depending on the potential depth. The left panel shows a shallow potential, the middle panel a one-phonon deep potential and the right panel a two-phonon deep potential. For the shallow potential the predominant desorption channel (bold orange) is a direct transition to the continuum, for a deep potential the cascade via the second level is the most important channel. Full lines are one-phonon processes, dashed lines two-phonon processes, and
$E_D=\hbar\omega_D=k_BT_D$.}
\label{2shuffle_sketch}
\end{figure}

The material parameters used for the numerical calculation apply to graphite, where the two-phonon
approximation is applicable. For a surface temperature of \(360~K\), we find an electron desorption time
\(2\cdot 10^{-5}~s\). Besides producing an estimate for the time \(\tau_{\rm e}\) with which an image-bound 
electron desorbs from a graphite surface, we also investigated, as a function of the surface temperature and the 
potential depth, the relative importance of direct vs. cascading desorption channels. For that purpose we 
used the Debye energy as an adjustable parameter.

As expected, the inverse desorption time, \(\tau_{\rm e}^{-1}\), depends strongly on the surface temperature, 
varying several orders of magnitude when the surface temperature changes. Depending on the depth of the surface 
potential we identified various desorption scenarios. They are summarized in Fig.~\ref{2shuffle_sketch}.
For a shallow potential, all transitions between the bound states and the continuum are one-phonon processes. The 
lowest bound state can be emptied directly to the continuum. This is more efficient than the detour via higher bound 
states. For a one-phonon deep potential the transition from the lowest bound state to the continuum is a two-phonon 
process, whereas the transition to the second bound state and from there to the continuum is a one-phonon 
process. In this case the cascade with two one-phonon processes is more efficient than the direct two-phonon process.
For a two-phonon deep potential both the direct transition from the lowest bound state to the continuum and the 
transition form the lowest bound state to the second bound state are two-phonon processes. The direct transition to 
the continuum is much slower, so that the detour via the second state is the faster channel. 

For most dielectrics of practical interest more than two phonons are required implying that for these materials
the desorption time for an image-bound electron may be in fact rather long. Indeed, in an ingenious experiment, 
using a field-effect transistor set-up with the gate replaced by an externally provided electron surface charge, 
Biasini and coworkers~\cite{BGY05,BGP05} determined the desorption time for an electron on a GaAs surface. They 
obtained $0.48~s$ which is rather long indeed but not unexpected, from our point of view,  because the energy 
difference between the lowest two image states of GaAs, obtained from the dynamically corrected image potential, 
is \(\Delta E_{12}=-0.152~eV\) implying more than 5 phonons to be necessary for that transition 
(\(\hbar\omega_D=0.03~eV\) for GaAs), which makes it accordingly unlikely. 

In principle, the device of Biasini and coworkers~\cite{BGY05,BGP05} would also allow to determine 
the electron sticking coefficient, making it a promising tool for a quantitative experimental investigation
of physisorption of electrons specifically at GaAs surfaces. The empirical data about $\tau_e$ and $s_e$ 
are however sparse in general. In view of the importance these two parameters have for the complete 
kinetic modeling of bounded plasmas, it is unacceptable to use them as adjustable parameters, as it 
is currently the case. We encourage therefore experimental groups to also design devices 
for the investigation of electron physisorption at surfaces which are used or naturally appear 
as boundaries of low-temperature gas discharges. 

%\begin{figure}
%\begin{minipage}{0.32\linewidth}
%\includegraphics[width=\linewidth]{2shuffle1.eps}
%\end{minipage}
%\begin{minipage}{0.32\linewidth}
%\includegraphics[width=\linewidth]{2shuffle2.eps}
%\end{minipage}
%\begin{minipage}{0.32\linewidth}
%\includegraphics[width=\linewidth]{2shuffle3.eps}
%\end{minipage}
%\caption{Desorption channels depending on the potential depth. The left panel shows a shallow potential, the middle panel a one-phonon deep potential and the right panel a two-phonon deep potential. For the shallow potential the predominant desorption channel (bold orange) is a direct transition to the continuum, for a deep potential the cascade via the second level is the most important channel. Full lines are one-phonon processes, dashed lines two-phonon processes.}
%\label{2shuffle_sketch}
%\end{figure}

%In this model scenario for desorption we have used a simple Debye spectrum. For a specific system a more realistic phonon spectrum had to be used. This would mollify the kinks and discontinuities marking the change of the potential depth, though we expect no qualitative change in the desorption behavior.

{\it Acknowledgments.}
This work was supported by the Deutsche Forschungsgemeinschaft (DFG) through the transregional collaborative 
research center SFB/TRR24. We acknowledge discussions with Hans Deutsch and Ralf Schneider.  

\appendix

\section{Schr\"odinger equation for surface states}
\label{schrss}

To obtain bound and unbound surface states the Schr\"odinger equation (\ref{sesufracestates}) has to be solved
with the proper boundary conditions. 

First, we consider bound states for which the wave functions have to vanish at \(x=x_c\) and for 
\(x\rightarrow\infty\). Substituting \(y=2\Lambda_0 x/\kappa\) with \(\kappa = \sqrt{- \Lambda_0^2 / \eta}\), 
Eq. (\ref{sesufracestates}) takes the form
\begin{align}
\phi^{\prime \prime}(y)+\left[\frac{\kappa}{y}-\frac{1}{4}\right]\phi(y)=0~,
\end{align}
whose solutions are Whittaker functions~\cite{WW27}. Hence, the wave functions which vanish at 
\(x=x_c\) and for \(x\rightarrow\infty\) are 
\begin{align}
\phi_q(x)=\frac{1}{\sqrt{a_B}\mathcal{N}_q} W_{\kappa,\frac{1}{2}}\left(\frac{2 \Lambda_0 x}{\kappa} \right),
\end{align}
where \(\mathcal{N}_q\) is a normalization constant defined by
\begin{align}
\mathcal{N}_q^2=\int_{x_c}^\infty \mathrm{d}x W_{\kappa,1/2}^2\left(\frac{2\Lambda_0 x}{\kappa}\right). 
\label{boundnormalisation}
\end{align}
The quantum number \(\kappa\) is determined by satisfying the boundary condition at the surface, 
\(W_{\kappa,1/2}(2 \Lambda_0 x_c/\kappa)=0\). This gives an infinite sequence of real numbers 
\(\kappa\) whose differences are however roughly one so that we can map them one-to-one onto 
integers \(q\), that is, \(q\leftrightarrow \kappa\). 
The energy of the bound state labelled with \(q\) is thus given by 
\begin{align}
E_q=-\frac{me^4 \Lambda_0^2}{2 \hbar^2}\frac{1}{\kappa^2}~.
\end{align}

Continuum states vanish only at \(x=x_c\). Since for them the energy is positive \(\kappa\) is imaginary. It 
is thus more convenient to label them with the real number \(k=1/(i\kappa)\). The wave function for 
continuum states is given by a linear combination of Whittaker functions,
\begin{align}
\phi_k(x)=\frac{1}{\mathcal{N}_k}\left[W_{-\frac{i}{k},\frac{1}{2}}(2i\Lambda_0 k x) + 
c W_{\frac{i}{k},\frac{1}{2}} (-2i\Lambda_0 k x)\right] \text{ ,} \label{continuumkwavefunction}
\end{align}
where the constant,
\begin{align}
c=-W_{-\frac{i}{k},\frac{1}{2}}(2i\Lambda_0 k x_c)/W_{\frac{i}{k},\frac{1}{2}} (-2i\Lambda_0 k x_c)~,
\end{align}
is chosen to enforce the boundary condition \(\phi(x_c)=0\). Normalizing the wave function in a box of 
length \(L\) leads to the normalization constant \( \mathcal{N}_k=\sqrt{2Le^{\pi/k}}\). 
The energy of the continuum states, finally, is given by
\begin{align}
E_k=\frac{me^4 \Lambda_0^2}{2 \hbar^2} k^2 \text{ .}
\end{align}
In the limit \(L\rightarrow\infty\) sums over continuum states can thus be transformed into integrals 
according to~\cite{Rafael09}
\begin{align}
\sum_{k>0} \dots=\frac{\Lambda_0 L}{a_B \pi}\int \mathrm{d}k \dots~,
\end{align}
where the \(L\) factor in front of the integral cancels with the \(\sqrt{L}\) factor contained in 
the normalization constant \(\mathcal{N}_k\) .

\section{Evaluation of the matrix elements}
\label{Approximations to the wave functions and matrix elements}

The electronic matrix element \(Z_{q,q^\prime}^n\) for two bound states labelled, respectively, by
\(q\) and \(q^\prime\), reads
\begin{align}
Z_{q,q^\prime}^n=\frac{1}{a_B^n \mathcal{N}_q\mathcal{N}_{q^\prime}}& \int_{x_c}^\infty \mathrm{d}x 
W_{\kappa,\frac{1}{2}}\left(\frac{2 \Lambda_0 x}{\kappa} \right) \nonumber \\ 
&\times  \frac{1}{x^n} W_{\kappa^\prime,\frac{1}{2}}\left(\frac{2\Lambda_0 x}{\kappa^\prime} \right) . 
\label{boundboundME}
\end{align}
For the efficient numerical evaluation of \(\mathcal{N}_q\) and \(Z_{q,q^\prime}^n\) we use an expansion 
of Whittaker functions for \(\kappa>0\) in terms of Laguerre polynomials~\cite{Rafael09}, 
\begin{align}
W_{\kappa,\frac{1}{2}}(x)=\sum_{n=0} \frac{\kappa (\kappa-1)e^{-\frac{1}{2}x}L_n(x) }
{(\kappa-n)(\kappa-n-1)\Gamma(2-\kappa)} \text{ .}
\end{align}

To calculate the bound state matrix element \(Z_{q,q^\prime}^n \) we can compute one matrix element after 
another. First, we  choose two states \(q\) and \(q^\prime\) which fixes the quantum numbers \(\kappa\) and 
\(\kappa^\prime\) in the bound state wave functions. Then, we integrate over \(x\) and obtain plain 
numbers for the matrix element which can be directly inserted into the calculation of the transition 
probabilities.

The evaluation of the electronic matrix element is more demanding if either one or both states are 
continuum states. A wave function in the continuum is labeled by a real number \(k\). If we were only 
interested in the value of the matrix element for some \(k\) we could follow the same strategy as for 
bound states. Some transition probabilities however contain sums over all electronic states, e.g., the sum over 
\(q_1\) in \(\mathcal{R}^{10}\) (see Eq. (\ref{examplerate10b})), which for continuum states implies an 
integral over \(k\). Hence \(k\) is not merely a parameter that we can specify in advance. It is 
rather a variable in a matrix element which thus becomes a function of \(k\),
\begin{align}
Z_{q,k}^n\rightarrow Z_q^n(k)=\int_{x_c}^\infty \mathrm{d}x \phi_{q}(x) 
\frac{1}{a_B^{n-1} x^n} \phi_k(x) \text{ .} 
\end{align}

For the two-phonon calculation we specifically need the bound-continuum matrix element \(Z_{q,k}^n\) 
for a given \(q\) and \(0<k<\infty\) and the continuum-continuum matrix element \(Z_{k,k^\prime}^2\) for 
\(0<k<\infty\) and \(k^\prime\) small. Because of the complicated structure of the Whittaker 
function the matrix elements cannot be obtained straightforwardly. To make their calculation 
feasible we constructed approximate expressions for the continuum wave function \(\phi_k(x)\) 
for small and large \(k\), respectively, calculated the matrix elements in these two limits, and 
then interpolated between them with a Pad\'{e} approximation~\cite{Rafael09}. 

We begin with the limit \(k\rightarrow 0\). In this limit, the Schr\"odinger equation for the continuum 
states is 
\begin{align}
\phi^{\prime\prime}(x)+\frac{2\Lambda_0}{x}\phi(x)=0 \text{ ,}
\end{align}
which after the substitutions \(t=2\sqrt{2\Lambda_0 x}\) and \(\phi= t \psi\) takes the form 
of the Bessel differential equation~\cite{MO48},
\begin{align}
\psi^{\prime \prime}+\frac{1}{t}\psi^\prime+\left(1-\frac{1}{t^2}\right)\psi=0~. \label{BesselsDE}
\end{align}
Hence, in the limit \(k\rightarrow 0\), the 
continuum wave function satisfying the boundary condition \(\phi_k(x_c)=0\) can be written 
as a linear combination of Bessel and Neumann functions.

There is however one technical caveat. Bessel and Neumann functions grow beyond limit for large \( x\) 
(see formula (9.2.1) in Ref.~\cite{AS73}) and cannot be normalized. Within the matrix 
element this is not dangerous because the decreasing factors \(1/x^n\) compensate 
the divergence at large \(x\). The only problem left is to find the prefactor by which
we have to multiply the linear combination so that it has for small \(x\) the amplitude 
of the correct \(\phi_{k\rightarrow 0}(x)\). Once we have this factor, the linear 
combination is normalized in the sense that its envelope coincides for small \(x\) with the 
envelope of the correct \(\phi_{k\rightarrow 0}(x)\). 

The most direct way to obtain the required multiplication factor is to perform the 
limit \(k\rightarrow 0\) in the continuum wave function (\ref{continuumkwavefunction}). 
Due to the complicated structure of the Whittaker function this is however not feasible.
Instead, it is better to determine the factor from the solution of Eq. (\ref{BesselsDE}),
\begin{align}
\phi(x)_{k\rightarrow 0}=2\sqrt{2\Lambda_0x} J_1(2\sqrt{2\Lambda_0 x}) \text{ ,} \label{Jk0sol}
\end{align}
which satisfies the boundary condition \(\phi(0)=0\). The solution of 
(\ref{sesufracestates}) satisfying the same boundary condition is given by 
\begin{align}
\phi_k(x)=\mathcal{\tilde{N}}_k^{-1} M_{-\frac{i}{k},\frac{1}{2}}(2i\Lambda_0 k x) \label{Mcontsol}
\end{align}
with the normalization constant 
\begin{align}
\mathcal{\tilde{N}}_k^{-1}=\sqrt{\frac{\pi}{Lk\left(1-e^{-2\pi / k}\right)}} \text{ .}  
\end{align}
In the limit \(k\rightarrow 0\) (\ref{Mcontsol}) merges into (\ref{Jk0sol}) because
\begin{align}
M_{-\frac{i}{k},\frac{1}{2}}(2i\Lambda_0 k x) \xrightarrow[k\rightarrow 0]{} 
i k\sqrt{2 \Lambda_0 x} J_1(\sqrt{8 \Lambda_0 x}) \text{ .}
\label{smallkxM}
\end{align} 
For the normalization constant we thus obtain \(\mathcal{N}_k^{-1}\rightarrow \sqrt{\pi/(Lk)}\) for \(k\rightarrow 0\).
%To find this normalization constant for the \(k\), we perform the limit \(k\rightarrow 0\) in the normalized continuum solution (\ref{continuumkwavefunction}). 
% It is however considerably simpler to use another continuum solution of (\ref{sesufracestates}) satisfying the boundary condition \(\phi(0)=0\), given by
%\begin{align}
%\phi_k(x)=\mathcal{\tilde{N}}_k^{-1} M_{-\frac{i}{k},\frac{1}{2}}(2i\Lambda_0 k x).
%\end{align}
%The limit \(k\rightarrow 0\) is given by
%\begin{align}
% M_{-\frac{i}{k},\frac{1}{2}}(2i\Lambda_0 k x) \xrightarrow[k\rightarrow 0]{} i k\sqrt{2 \Lambda_0 x} J_1(\sqrt{8 \Lambda_0 x}) \text{ ,}
%\label{smallkxM}
%\end{align}
%which leads to
%\begin{align}
%\mathcal{\tilde{N}}_k^{-1}=\sqrt{\frac{\pi}{Lk\left(1-e^{-2\pi / k}\right)}} \xrightarrow[k\rightarrow 0]{} \sqrt{\frac{\pi}{Lk}} \text{ .}  
%\end{align}
The fact that \(k\) and \(x\) are tied together in the argument of \(M_{-i/k,1/2}(2i\Lambda_0 k x)\) suggests 
that an approximation for small \(k\) is at the same time an approximation for small \(x\). Since we need the 
wave function only for small \(x\) and small \(k\), we expect the replacement (\ref{smallkxM}) 
to provide good results for the desired normalization factor. 

Since the amplitudes of Bessel and Neumann functions are the same for large \(x\), we can use the normalization 
obtained for the Bessel function for the Neumann function as well. A normalized approximation satisfying the 
correct boundary condition, \(\phi_k(x_c)=0\), is then given by
\begin{align}
\phi(x)_{k\rightarrow 0} =& \frac{1}{\sqrt{L}}  \sqrt{\frac{\pi}{1+\tilde{c}^2}}   \sqrt{k 2\Lambda_0 x} \nonumber \\
 &\quad \times \left[J_1(2\sqrt{2\Lambda_0 x}) - \tilde{c} N_1(2\sqrt{2\Lambda_0 x})\right] \label{xcbck0sol}
\end{align} 
with
\begin{align}
\tilde{c} =-J_1(2\sqrt{2\Lambda_0 x_c})/N_1(2\sqrt{2\Lambda_0 x_c})~.   
\end{align}

Using (\ref{xcbck0sol}) the matrix element \(Z_{q k}^n\) can be evaluated in the limit \(k\rightarrow 0\). We find
\begin{align}
Z_{q,k\rightarrow 0}^n=\int_{x_c}^\infty\!\!\!\!dx\frac{\phi_q(x)\phi_{k\rightarrow 0}(x)}{a_B^{n-1}x^n}
=\frac{\alpha_q^n\sqrt{k}}{a_B^{n-1/2}\sqrt{L}\mathcal{N}_q}~,
\end{align}
where the \(k\) dependency is separated so that the remaining integral gives a \(k\)-independent quantity, 
\(\alpha_q^n\), which has to be obtained numerically. Similarly,
\begin{align}
Z_{k\rightarrow 0,k^\prime \rightarrow 0}=\int_{x_c}^\infty\!\!\!\!dx 
\frac{\phi_{k^\prime \rightarrow 0}(x)\phi_{k\rightarrow 0}(x)}{a_B x^2}=\frac{\alpha_c^2\sqrt{kk^\prime}}{L a_B}
\end{align}
with \(\alpha_c^2\) again a constant to be determined numerically.

We now proceed to the approximation of the continuum wave functions for large \(k\). The higher the energy of 
the continuum states, the lesser the potential at the surface changes the plane wave behavior far from the
surface. Therefore, in the limit \(k\rightarrow \infty\) we can accurately describe the wave function by a plane 
wave. Using the asymptotic form for the Whittaker function (page 116 in Ref.~\cite{MO48}),
\begin{align}
W_{-\frac{i}{k},\frac{1}{2}}(2i\Lambda_0 k x) &\approx e^{-i\Lambda_0 k x} e^{\frac{\pi}{2k}}~,
\end{align}
the continuum wave function in the limit \(k\rightarrow \infty\) satisfying the correct boundary condition,
\(\phi_k(x_c)=0\), can be approximated by 
\begin{align}
\phi_k(x)=\sqrt{2/L} \sin(\Lambda_0 k (x-x_c)) \text{ .} \label{contlargekwavefct}
\end{align}

Employing, finally, the Fourier integral~\cite{Olver74} the matrix elements for large \(k\) can be shown 
to be~\cite{Rafael09}
\begin{align}
Z_{q,k\rightarrow \infty}^n&=\frac{1}{a_B^{n-1/2}\sqrt{L} \mathcal{N}_q}
\frac{2\sqrt{2}n}{(\Lambda_0 k)^3 x_c^{n+1}}W_{\kappa,\frac{1}{2}}^\prime 
\left(\frac{2 \Lambda_0}{\kappa} x_c \right) \nonumber \\
&=\frac{1}{a_B^{n-1/2}\sqrt{L}\mathcal{N}_q}b_q^n \frac{1}{k^3}
\end{align}
and 
\begin{align}
Z_{k^\prime \rightarrow 0,k\rightarrow \infty}^2=&\frac{1}{L a_B} \sqrt{\frac{\pi}{1+\tilde{c}^2}} \frac{32}{\Lambda_0^{3/2}x_c^3} \nonumber \\
& \times \left[J_0(\sqrt{8\Lambda_0 x_c})-\tilde{c} N_0(\sqrt{8 \Lambda_0 x_c}) \right] \frac{\sqrt{k^\prime}}{k^3} \nonumber \\
=&\frac{1}{L a_B} b_c^2 \frac{\sqrt{k^\prime}}{k^3}~,
\end{align}
where the \(k-{\rm independent}\) coefficients \(b_q^n\) and \(b_c^2\) have to be worked out again numerically.

Having calculated the leading terms for the matrix elements for small and large \(k\), we can combine these 
two limits via a Pad\'{e} approximation in terms of \(\sqrt{k}\). The coefficients are chosen in such a way 
that the Pad\'{e} approximation matches the leading term of both limits: \(k\rightarrow 0\) and 
\(k \rightarrow \infty\). Then, the matrix elements read
\begin{align}
Z_{q,k}^n = \frac{1}{\sqrt{L}  a_B^{n-1/2} \mathcal{N}_q} \frac{\alpha_q^n k^{1/2}}{1+\beta_q^n k^{7/2}} \text{ ,} 
\label{boundstatecontinuumelectronicme}
\end{align}
where \(\beta_q^n=\alpha_q^n / b_q^n\), and
\begin{align}
Z_{k,k^\prime}^2 = \frac{1}{L a_B} \frac{\alpha_{c}^2 k^{\prime 1/2} k^{1/2}}{1+\beta_{c}^2 k^{7/2}} \text{ ,}
\label{contcontME}
\end{align}
where \(\beta_c^2=\alpha_c^2 /b_c^2\).\\ 

\section{Transition probabilities in a compact form}
\label{compactrates}

In this appendix we list the one- and two-phonon transition probabilities as used in the numerical calculation of 
the desorption time. We implicitly assume that $q$ labels both bound and unbound states. For a bound state 
$q$ is simply an integer (to be mapped onto $\kappa(q)$) whereas for a continuum state $q$ stands for a real 
number $k$. To obtain a compact form for the transition probabilities it is moreover convenient to introduce 
dimensionless variables, 
\begin{align}
x&=\frac{\omega}{\omega_D}~,\\ 
\quad \epsilon_q&=\frac{E_q}{\hbar \omega_D}~,\\ 
\quad \Delta_{q,q^\prime}&=\frac{E_q-E_{q^\prime}}{\hbar \omega_D}~, \\
\quad \delta&=\frac{\hbar \omega_D}{k_B T}~, \\ 
\quad \nu(x)&=\frac{\gamma_{\omega}}{\omega_D} \text{ .}
\end{align}

The transition probability of \(\mathcal{O}(u) \), that is, the one-phonon transition probability employed in a 
golden rule approximation, is given by
\begin{widetext}
\begin{align}
\mathcal{R}^{1}\left(q^{\prime},q \right)&=\frac{2\pi}{\hbar} \sum_{Q}   G_{q,q^{\prime}}^{1} \left(Q\right)  \left[G_{q,q^{\prime}}^{1}\left(Q\right)\right]^\ast \big( n_{B}\left(\hbar \omega_{Q} \right) \delta \left(E_q-E_{q^{\prime}}+\hbar \omega_Q \big)
%  \right. \nonumber \\
%& \left.  
+  \left[ 1+ n_{B}\left(\hbar \omega_{Q} \right)\right]  \delta \left(E_q-E_{q^{\prime}}-\hbar \omega_Q \right)  \right) \nonumber  \\
&=\frac{3 \pi e^4 \Lambda_0^2}{\hbar \mu \omega_{D}^{2}}  Z_{q,q^{\prime}}^{2} Z_{q,q^{\prime}}^{2} \left(I_{(1)}^{1}(q^\prime,q)+I_{(1)}^{2}(q^\prime,q) \right), \label{onephononcompactrate}
\end{align}
where 
\begin{align}
I_{(1)}^{1}(q^\prime,q)&=\frac{-\Delta_{q,q^\prime}}{e^{-\delta \Delta_{q,q^\prime}}-1}  ~\textnormal{ for }~ 0\le-\Delta_{q,q^\prime}\le 1 \text{ ,}
\end{align}
and
\begin{align}
I_{(1)}^{2}(q^\prime,q)&=\Delta_{q,q^\prime} \left[1+\frac{1}{e^{\delta \Delta_{q,q^\prime}}-1} \right]  ~\textnormal{ for }~ 0\le\Delta_{q,q^\prime}\le 1  \text{ .}
\end{align}
Otherwise \(I_{(1)}^1\) and \(I_{(1)}^2\) are zero. Depending on whether $q$ and $q'$ denote bound or 
continuum states, the electronic matrix element \(Z_{q,q^{\prime}}^{2}\) is either given by (\ref{boundboundME}), 
(\ref{boundstatecontinuumelectronicme}), or (\ref{contcontME}). 

%\subparagraph{Unregularized two-phonon rates.} 
\noindent
As explained in Subsection \ref{oneANDtwo}, we keep only those parts of the transition probabilities 
of \(\mathcal{O}(u^{4}) \)
which give rise to what we call two-Debye-energy transitions, which are transitions between states that 
are between one and two Debye energies apart. In the unregularized form, that is, in the form which 
diverges in particular situations (see Subsection \ref{Regularization} for a discussion), the two-phonon 
transition probabilities included in our calculation are given by
%The rate \(\mathcal{R}^9(q^\prime,q)\): 
\begin{align}
\mathcal{\tilde{R}}^{9}\left(q^{\prime},q \right)&=\frac{2\pi}{\hbar} 
\sum_{Q_1,Q_2} G_{q^\prime ,q}^{2}\left(Q_{1},Q_1\right)
\left[G_{q^\prime ,q}^{2}\left(Q_{2},Q_2\right)\right]^\ast  
\left(2 n_{B}\left(\hbar \omega_{Q_1} \right) n_{B}\left(\hbar \omega_{Q_2} \right) 
\delta\left(E_q-E_{q^\prime}+\hbar \omega_{Q_1}+\hbar \omega_{Q_2} \right) \right.\nonumber \\
&\left. 
+ 2 \left[1+ n_{B}\left(\hbar \omega_{Q_1} \right)\right]\left[1+ n_{B}\left(\hbar \omega_{Q_2} \right)\right] 
\delta\left(E_q-E_{q^\prime}-\hbar \omega_{Q_1}-\hbar \omega_{Q_2} \right)_{ }^{ } \right) \nonumber \\
&= \frac{9 \pi e^4 \Lambda_0^2}{ \mu^2 \omega_{D}^{3}} Z_{q,q^{\prime}}^{3} Z_{q,q^{\prime}}^{3} \left(I_{(2)}^{1}(q^\prime,q)+I_{(2)}^{2}(q^\prime,q) \right) \text{ ,}\label{compactr9}\\
%\end{align}
%The rate \(\mathcal{R}^{10}(q^\prime,q)\): 
%\begin{align}
\mathcal{\tilde{R}}^{10}\left(q^{\prime},q \right)&=-\frac{2\pi}{\hbar} \sum_{Q_1,Q_2}  G_{q^\prime ,q}^{2}\left(Q_{1},Q_1\right)  \left[G_{q_{1},q}^{1}\left(Q_{2}\right)    G_{q^\prime ,q_1}^{1}\left(Q_{2}\right) \right]^\ast \nonumber \\
&\times \left( 2   n_{B}\left(\hbar \omega_{Q_1} \right)  n_{B}\left(\hbar \omega_{Q_2} \right) \frac{\delta\left(E_{q}-E_{q^\prime}+\hbar\omega_{Q_{1}}+\hbar\omega_{Q_{2}} \right)}{E_q-E_{q_1}+\hbar\omega_{Q_1}-i\epsilon}   \right.  \nonumber  \\
&\left. + 2\left[1+ n_{B}\left(\hbar \omega_{Q_1} \right)\right] \left[1+ n_{B}\left(\hbar \omega_{Q_2} \right)\right] \frac{\delta\left(E_{q}-E_{q^\prime}-\hbar\omega_{Q_{1}}-\hbar\omega_{Q_{2}} \right)}{E_q-E_{q_1}-\hbar\omega_{Q_1}-i\epsilon} \right)  \nonumber  \\
&=-\frac{9\pi e^6 \Lambda_0^3}{\hbar \mu^2 \omega_D^4} \sum_{q_1} Z_{q^\prime,q}^{3} Z_{q,q_1}^2 Z_{q_1,q^\prime}^2 \left(I_{(2)}^3 \left(q^\prime,q;q_1\right) + I_{(2)}^4 \left(q^\prime,q;q_1\right) \right) \text{ ,}  \label{compactr10}\\ 
%\end{align}
%The rate \(\mathcal{R}^{14}(q^\prime,q)\): 
%\begin{align}
\mathcal{\tilde{R}}^{14}\left(q^{\prime},q \right)&=\frac{2\pi}{\hbar} \sum_{q_1,q_2} \sum_{Q_1,Q_2} G_{q^\prime ,q_1}^{1}\left(Q_{1}\right)  G_{q_1 ,q}^{1}\left(Q_{1}\right)  \left[G_{q^\prime ,q_2}^{1}\left(Q_{2}\right)  G_{q_2 ,q}^{1}\left(Q_{2}\right) \right]^\ast \nonumber \\
& \times \left(  n_{B}\left(\hbar \omega_{Q_1} \right)  n_{B}\left(\hbar \omega_{Q_2} \right) 
\frac{\delta\left(E_{q}-E_{q^\prime}+\hbar\omega_{Q_{1}}+\hbar\omega_{Q_{2}} \right)}{(E_q-E_{q_1}+\hbar\omega_{Q_2}+i\epsilon)(E_q-E_{q_2}+\hbar\omega_{Q_1}-i\epsilon)}  \right. \nonumber \\
& \left. + n_{B}\left(\hbar \omega_{Q_1} \right)  n_{B}\left(\hbar \omega_{Q_2} \right) 
\frac{\delta\left(E_{q}-E_{q^\prime}+\hbar\omega_{Q_{1}}+\hbar\omega_{Q_{2}} \right)}
{(E_q-E_{q_1}+\hbar\omega_{Q_1}+i\epsilon)(E_q-E_{q_2}+\hbar\omega_{Q_1}-i\epsilon)}  \right. \nonumber \\
& \left.+ \left[1+ n_{B}\left(\hbar \omega_{Q_1} \right)\right] \left[1+ n_{B}\left(\hbar \omega_{Q_2} \right)\right] 
\frac{\delta\left(E_{q}-E_{q^\prime}-\hbar\omega_{Q_{1}}-\hbar\omega_{Q_{2}} \right)}
{(E_q-E_{q_1}-\hbar\omega_{Q_2}+i\epsilon)(E_q-E_{q_2}-\hbar\omega_{Q_1}-i\epsilon)} \right. \nonumber \\
& \left. + \left[1+ n_{B}\left(\hbar \omega_{Q_1} \right)\right] \left[1+ n_{B}\left(\hbar \omega_{Q_2} \right)\right] 
\frac{\delta\left(E_{q}-E_{q^\prime}-\hbar\omega_{Q_{1}}-\hbar\omega_{Q_{2}} \right)}
{(E_q-E_{q_1}-\hbar\omega_{Q_2}+i\epsilon)(E_q-E_{q_2}-\hbar\omega_{Q_2}-i\epsilon)}  \right) \nonumber \\
&=\frac{9\pi e^8 \Lambda_0^4}{2 \hbar^2 \mu^2 \omega_D^5} \sum_{q_1,q_2} Z_{q^\prime,q_1}^{2}  Z_{q_1,q}^{2}  Z_{q,q_2}^{2}  Z_{q_2,q^\prime}^{2} \left(I_{(2)}^5 \left(q^\prime,q;q_1,q_2\right) +I_{(2)}^6 \left(q^\prime,q;q_1,q_2\right) +I_{(2)}^7 \left(q^\prime,q;q_1,q_2\right) \right. \nonumber\\ 
& \left. + I_{(2)}^8 \left(q^\prime,q;q_1,q_2\right)\right) \text{ ,}\label{compactr14}
\end{align}
where, in the limit \(\epsilon \rightarrow 0 \), the auxiliary integrals are defined by 
\begin{flalign}
\begin{split}
I_{(2)}^1(q^\prime,q)&=\int_{-\Delta_{q,q^\prime}-1}^1 \mathrm{d}x \frac{x}{e^{\delta x}-1} \frac{-\Delta_{q,q^\prime}-x}{e^{\delta \left( - \Delta_{q,q^\prime} -x\right)}- 1} \text{ ,}
\end{split} & \\
\begin{split}
I_{(2)}^2(q^\prime,q)&=\int_{\Delta_{q,q^\prime}-1}^1 \mathrm{d}x x\left[1+ \frac{1}{e^{\delta x}-1} \right] \left( \Delta_{q,q^\prime}-x \right) \left[1+\frac{1}{e^{\delta \left( \Delta_{q,q^\prime} -x \right)}-1}\right]\text{ ,}
\end{split} &
\end{flalign}
\begin{flalign}
\begin{split}
I_{(2)}^3\left(q^\prime,q;q_1\right)&=\int_{-\Delta_{q,q^\prime}-1}^1 \mathrm{d}x \frac{x}{e^{\delta x}-1} \frac{-\Delta_{q,q^\prime}-x}{e^{\delta \left( - \Delta_{q,q^\prime} -x\right)}- 1} \frac{1}{\Delta_{q,q_1}+x-i\epsilon} \text{ ,} \label{I3def}\end{split} & \\
\begin{split}
I_{(2)}^4\left(q^\prime,q;q_1\right)&=\int_{\Delta_{q,q^\prime}-1}^1 \mathrm{d}x x\left[1+ \frac{1}{e^{\delta x}-1} \right] \left( \Delta_{q,q^\prime}-x \right) \left[1+\frac{1}{e^{\delta \left( \Delta_{q,q^\prime} -x \right)}-1}\right] \frac{1}{\Delta_{q,q_1}-x-i\epsilon} \text{ ,}
\end{split} &
\end{flalign}
\begin{flalign}
\begin{split}
I_{(2)}^5\left(q^\prime,q;q_1,q_2\right)&= \int_{-\Delta_{q,q^\prime}-1}^1 \mathrm{d}x \frac{x}{e^{\delta x}-1} \frac{-\Delta_{q,q^\prime}-x}{e^{\delta \left( - \Delta_{q,q^\prime} -x\right)}- 1}  \frac{1}{\Delta_{q^\prime ,q_1}-x+i\epsilon}  \frac{1}{\Delta_{q,q_2}+x-i\epsilon} \text{ ,}\end{split} & \\
\begin{split}
I_{(2)}^6\left(q^\prime,q;q_1,q_2\right)&=  \int_{-\Delta_{q,q^\prime}-1}^1 \mathrm{d}x \frac{x}{e^{\delta x}-1} \frac{-\Delta_{q,q^\prime}-x}{e^{\delta \left( - \Delta_{q,q^\prime} -x\right)}- 1}  \frac{1}{\Delta_{q ,q_1}+x+i\epsilon}  \frac{1}{\Delta_{q,q_2}+x-i\epsilon}\text{ ,}\label{I6def}
\end{split} &
\end{flalign}
\begin{flalign}
\begin{split}
I_{(2)}^7\left(q^\prime,q;q_1,q_2\right)&= \int_{\Delta_{q,q^\prime}-1}^1 \mathrm{d}x x\left[1+ \frac{1}{e^{\delta x}-1} \right] \left( \Delta_{q,q^\prime}-x \right) \left[1+\frac{1}{e^{\delta \left( \Delta_{q,q^\prime} -x \right)}-1}\right]\frac{1}{\Delta_{q^\prime ,q_1}+x+i\epsilon} \frac{1}{\Delta_{q,q_2}-x-i\epsilon}\text{ ,} \end{split} & \\
\begin{split}
I_{(2)}^8\left(q^\prime,q;q_1,q_2\right)&=  \int_{\Delta_{q,q^\prime}-1}^1 \mathrm{d}x x\left[1+ \frac{1}{e^{\delta x}-1} \right] \left( \Delta_{q,q^\prime}-x \right) \left[1+\frac{1}{e^{\delta \left( \Delta_{q,q^\prime} -x \right)}-1}\right]  \frac{1}{\Delta_{q ,q_1}-x+i\epsilon} \frac{1}{\Delta_{q,q_2}-x-i\epsilon}\text{ .}
\end{split} &
\end{flalign}

%\subparagraph{Regularized two-phonon rates.}
\noindent
The two-phonon transition probabilities \(\mathcal{\tilde{R}}^{10}\) and \(\mathcal{\tilde{R}}^{14}\) 
diverge for the situations studied in this paper.  They have to be therefore regularized. 
The regularization procedure outlined 
in Subsection \ref{Regularization} leads then to the divergence-free, regularized transition probabilities,
%\(\mathcal{R}^{10}(q^\prime,q)\):
\begin{align}
\mathcal{\tilde{R}}^{10}(q^\prime,q)=-\frac{9\pi e^6 \Lambda_0^3}{\hbar \mu^2 \omega_D^4} \sum_{q_1} Z_{q^\prime,q}^3 Z_{q^\prime,q_1}^2  Z_{q_1,q}^2 \left(I_{(2)}^{3^\prime}(q^\prime,q;q_1)+I_{(2)}^{4^\prime}(q^\prime,q;q_1) \right)\text{ ,} \label{r10compactcorr}
\end{align}
%The rate \(\mathcal{R}^{14}(q^\prime,q)\):
\begin{align}
\mathcal{\tilde{R}}^{14}(q^\prime,q)=\frac{9\pi e^8 \Lambda_0^4}{2 \hbar^2 \mu^2 \omega_D^5} \sum_{q_1,q_2} Z_{q^\prime ,q_1}^{2}  Z_{q_1 ,q}^{2}  Z_{q ,q_2}^{2}  Z_{q_2 ,q^\prime}^{2} \left(I_{(2)}^{5^\prime} \left(q^\prime,q;q_1,q_2\right)  +I_{(2)}^{7^\prime} \left(q^\prime,q;q_1,q_2\right) \right)\text{ ,} \label{r14compactcorr}
\end{align}
where the integrals are given by 
\begin{flalign}
\begin{split}
I_{(2)}^{3^\prime}(q^\prime,q;q_1)=&\int_{-\Delta_{q,q^\prime}-1}^1 \!\!\!\!\mathrm{d}x \frac{x}{e^{\delta x}-1} \frac{-\Delta_{q,q^\prime}-x}{e^{\delta \left( - \Delta_{q,q^\prime} -x\right)}- 1} g(\Delta_{q,q_1}+x,\nu(-x-\Delta_{q,q^\prime})+2\nu(x)) \text{ ,}\end{split} & \\
\begin{split}
I_{(2)}^{4^\prime}\left(q^\prime,q;q_1\right)=&\int_{\Delta_{q,q^\prime}-1}^1 \!\!\!\!\!\!\!\mathrm{d}x x\left[1+ \frac{1}{e^{\delta x}-1} \right] \left( \Delta_{q,q^\prime}-x \right) \left[1+\frac{1}{e^{\delta \left( \Delta_{q,q^\prime} -x \right)}-1}\right]  g(\Delta_{q,q_1}-x,\nu(\Delta_{q,q^\prime}-x)+2\nu(x))\text{ ,}
\end{split} &
\end{flalign}
\begin{flalign}
\begin{split}
I_{(2)}^{5^\prime} \left(q^\prime,q;q_1,q_2\right)&=\int_{-\Delta_{q,q^\prime}-1}^1 \mathrm{d}x \frac{x}{e^{\delta x}-1} \frac{-\Delta_{q,q^\prime}-x}{e^{\delta \left( - \Delta_{q,q^\prime} -x\right)}- 1}   \\
& \times \left[f(\Delta_{q^\prime,q_1}-x,\Delta_{q,q_2}+x,\nu(x)+2\nu(-x-\Delta_{q,q^\prime}),2\nu(x)+\nu(-x-\Delta_{q,q^\prime})) 
\right.  \\
&\left. \quad+ f(\Delta_{q,q_1}+x,\Delta_{q,q_2}+x,2\nu(x)+\nu(-x-\Delta_{q,q^\prime}),2\nu(x)+\nu(-x-\Delta_{q,q^\prime})) \right] \text{ ,}
\end{split}&
\end{flalign}
\begin{flalign}
\begin{split}
I_{(2)}^{7^\prime} \left(q^\prime,q;q_1,q_2\right)=& \int_{\Delta_{q,q^\prime}-1}^1 \mathrm{d}x x\left[1+ \frac{1}{e^{\delta x}-1} \right] \left( \Delta_{q,q^\prime}-x \right) \left[1+\frac{1}{e^{\delta \left( \Delta_{q,q^\prime} -x \right)}-1}\right]   \\
& \times \left[ f(\Delta_{q^\prime,q_1}+x,\Delta_{q,q_2}-x,\nu(x)+2\nu(\Delta_{q,q^\prime}-x),2\nu(x)+\nu(\Delta_{q,q^\prime}-x)) \right.   \\
& \left. \quad+f(\Delta_{q,q_1}-x,\Delta_{q,q_2}-x,2\nu(x)+\nu(\Delta_{q,q^\prime}-x),2\nu(x)+\nu(\Delta_{q,q^\prime}-x)) \right] \text{ .}
\end{split}&
\end{flalign}

\end{widetext}


\begin{thebibliography}{66}
\expandafter\ifx\csname natexlab\endcsname\relax\def\natexlab#1{#1}\fi
\expandafter\ifx\csname bibnamefont\endcsname\relax
  \def\bibnamefont#1{#1}\fi
\expandafter\ifx\csname bibfnamefont\endcsname\relax
  \def\bibfnamefont#1{#1}\fi
\expandafter\ifx\csname citenamefont\endcsname\relax
  \def\citenamefont#1{#1}\fi
\expandafter\ifx\csname url\endcsname\relax
  \def\url#1{\texttt{#1}}\fi
\expandafter\ifx\csname urlprefix\endcsname\relax\def\urlprefix{URL }\fi
\providecommand{\bibinfo}[2]{#2}
\providecommand{\eprint}[2][]{\url{#2}}

\bibitem[{\citenamefont{Boettcher}(1952)}]{Boettcher52}
\bibinfo{author}{\bibfnamefont{C.~J.~F.} \bibnamefont{Boettcher}},
  \emph{\bibinfo{title}{Theory of electric polarization}}
  (\bibinfo{publisher}{Elsevier Publishing Company},
  \bibinfo{address}{Amsterdam}, \bibinfo{year}{1952}).

\bibitem[{\citenamefont{Cole and Cohen}(1969)}]{CC69}
\bibinfo{author}{\bibfnamefont{M.~W.} \bibnamefont{Cole}} \bibnamefont{and}
  \bibinfo{author}{\bibfnamefont{M.~H.} \bibnamefont{Cohen}},
  \bibinfo{journal}{Phys. Rev. Lett.} \textbf{\bibinfo{volume}{23}},
  \bibinfo{pages}{1238} (\bibinfo{year}{1969}).

\bibitem[{\citenamefont{Dose et~al.}(1984)\citenamefont{Dose, Altmann,
  Goldmann, Kolac, and Rogozik}}]{DAG84}
\bibinfo{author}{\bibfnamefont{V.}~\bibnamefont{Dose}},
  \bibinfo{author}{\bibfnamefont{W.}~\bibnamefont{Altmann}},
  \bibinfo{author}{\bibfnamefont{A.}~\bibnamefont{Goldmann}},
  \bibinfo{author}{\bibfnamefont{U.}~\bibnamefont{Kolac}}, \bibnamefont{and}
  \bibinfo{author}{\bibfnamefont{J.}~\bibnamefont{Rogozik}},
  \bibinfo{journal}{Phys. Rev. Lett.} \textbf{\bibinfo{volume}{52}},
  \bibinfo{pages}{1919} (\bibinfo{year}{1984}).

\bibitem[{\citenamefont{Straub and Himpsel}(1984)}]{SH84}
\bibinfo{author}{\bibfnamefont{D.}~\bibnamefont{Straub}} \bibnamefont{and}
  \bibinfo{author}{\bibfnamefont{F.~J.} \bibnamefont{Himpsel}},
  \bibinfo{journal}{Phys. Rev. Lett.} \textbf{\bibinfo{volume}{52}},
  \bibinfo{pages}{1922} (\bibinfo{year}{1984}).

\bibitem[{\citenamefont{Woodruff et~al.}(1985)\citenamefont{Woodruff, Hulbert,
  Johnson, and Smith}}]{WHJ85}
\bibinfo{author}{\bibfnamefont{D.~P.} \bibnamefont{Woodruff}},
  \bibinfo{author}{\bibfnamefont{S.~L.} \bibnamefont{Hulbert}},
  \bibinfo{author}{\bibfnamefont{P.~D.} \bibnamefont{Johnson}},
  \bibnamefont{and} \bibinfo{author}{\bibfnamefont{N.~V.} \bibnamefont{Smith}},
  \bibinfo{journal}{Phys. Rev. B} \textbf{\bibinfo{volume}{31}},
  \bibinfo{pages}{(RC)4046} (\bibinfo{year}{1985}).

\bibitem[{\citenamefont{Jacob et~al.}(1986)\citenamefont{Jacob, Dose, Kolac,
  and Fauster}}]{JDK86}
\bibinfo{author}{\bibfnamefont{W.}~\bibnamefont{Jacob}},
  \bibinfo{author}{\bibfnamefont{V.}~\bibnamefont{Dose}},
  \bibinfo{author}{\bibfnamefont{U.}~\bibnamefont{Kolac}}, \bibnamefont{and}
  \bibinfo{author}{\bibfnamefont{T.}~\bibnamefont{Fauster}},
  \bibinfo{journal}{Z. Phys. B} \textbf{\bibinfo{volume}{63}},
  \bibinfo{pages}{459} (\bibinfo{year}{1986}).

\bibitem[{\citenamefont{Fauster}(1994)}]{Fauster94}
\bibinfo{author}{\bibfnamefont{T.}~\bibnamefont{Fauster}},
  \bibinfo{journal}{Appl. Phys. A} \textbf{\bibinfo{volume}{59}},
  \bibinfo{pages}{479} (\bibinfo{year}{1994}).

\bibitem[{\citenamefont{Hoefer et~al.}(1997)\citenamefont{Hoefer, Shumay,
  Reuss, Thomann, Wallauer, and Fauster}}]{HSR97}
\bibinfo{author}{\bibfnamefont{U.}~\bibnamefont{Hoefer}},
  \bibinfo{author}{\bibfnamefont{I.~L.} \bibnamefont{Shumay}},
  \bibinfo{author}{\bibfnamefont{C.}~\bibnamefont{Reuss}},
  \bibinfo{author}{\bibfnamefont{U.}~\bibnamefont{Thomann}},
  \bibinfo{author}{\bibfnamefont{W.}~\bibnamefont{Wallauer}}, \bibnamefont{and}
  \bibinfo{author}{\bibfnamefont{T.}~\bibnamefont{Fauster}},
  \bibinfo{journal}{Science} \textbf{\bibinfo{volume}{277}},
  \bibinfo{pages}{1480} (\bibinfo{year}{1997}).

\bibitem[{\citenamefont{Hoefer}(1999)}]{Hoefer99}
\bibinfo{author}{\bibfnamefont{U.}~\bibnamefont{Hoefer}},
  \bibinfo{journal}{Appl. Phys. B} \textbf{\bibinfo{volume}{68}},
  \bibinfo{pages}{383} (\bibinfo{year}{1999}).

\bibitem[{\citenamefont{Fauster and Weinelt}(2005)}]{FW05}
\bibinfo{author}{\bibfnamefont{T.}~\bibnamefont{Fauster}} \bibnamefont{and}
  \bibinfo{author}{\bibfnamefont{M.}~\bibnamefont{Weinelt}},
  \bibinfo{journal}{Surface science} \textbf{\bibinfo{volume}{593}},
  \bibinfo{pages}{1} (\bibinfo{year}{2005}).

\bibitem[{\citenamefont{Gumhalter et~al.}(2008)\citenamefont{Gumhalter,
  \v{S}iber, Buljan, and Fauster}}]{GSB08}
\bibinfo{author}{\bibfnamefont{B.}~\bibnamefont{Gumhalter}},
  \bibinfo{author}{\bibfnamefont{A.}~\bibnamefont{\v{S}iber}},
  \bibinfo{author}{\bibfnamefont{H.}~\bibnamefont{Buljan}}, \bibnamefont{and}
  \bibinfo{author}{\bibfnamefont{T.}~\bibnamefont{Fauster}},
  \bibinfo{journal}{Phys. Rev. B} \textbf{\bibinfo{volume}{78}},
  \bibinfo{pages}{155410} (\bibinfo{year}{2008}).

\bibitem[{\citenamefont{Cole}(1974)}]{Cole74}
\bibinfo{author}{\bibfnamefont{M.~W.} \bibnamefont{Cole}},
  \bibinfo{journal}{Rev. Mod. Phys.} \textbf{\bibinfo{volume}{46}},
  \bibinfo{pages}{451} (\bibinfo{year}{1974}).

\bibitem[{\citenamefont{Lehmann et~al.}(1999)\citenamefont{Lehmann, Merschdorf,
  Thon, Voll, and Pfeiffer}}]{LMT99}
\bibinfo{author}{\bibfnamefont{J.}~\bibnamefont{Lehmann}},
  \bibinfo{author}{\bibfnamefont{M.}~\bibnamefont{Merschdorf}},
  \bibinfo{author}{\bibfnamefont{A.}~\bibnamefont{Thon}},
  \bibinfo{author}{\bibfnamefont{S.}~\bibnamefont{Voll}}, \bibnamefont{and}
  \bibinfo{author}{\bibfnamefont{W.}~\bibnamefont{Pfeiffer}},
  \bibinfo{journal}{Phys. Rev. B} \textbf{\bibinfo{volume}{60}},
  \bibinfo{pages}{17037} (\bibinfo{year}{1999}).

\bibitem[{\citenamefont{Kutschera et~al.}(2007)\citenamefont{Kutschera,
  Weinelt, Rohlfing, and Fauster}}]{KWR07}
\bibinfo{author}{\bibfnamefont{M.}~\bibnamefont{Kutschera}},
  \bibinfo{author}{\bibfnamefont{M.}~\bibnamefont{Weinelt}},
  \bibinfo{author}{\bibfnamefont{M.}~\bibnamefont{Rohlfing}}, \bibnamefont{and}
  \bibinfo{author}{\bibfnamefont{T.}~\bibnamefont{Fauster}},
  \bibinfo{journal}{Appl. Phys. A} \textbf{\bibinfo{volume}{88}},
  \bibinfo{pages}{519} (\bibinfo{year}{2007}).

\bibitem[{\citenamefont{Himpsel et~al.}(1979)\citenamefont{Himpsel, Knapp,
  VanVechten, and Eastman}}]{HKV79}
\bibinfo{author}{\bibfnamefont{F.~J.} \bibnamefont{Himpsel}},
  \bibinfo{author}{\bibfnamefont{J.~A.} \bibnamefont{Knapp}},
  \bibinfo{author}{\bibfnamefont{J.~A.} \bibnamefont{VanVechten}},
  \bibnamefont{and} \bibinfo{author}{\bibfnamefont{D.~E.}
  \bibnamefont{Eastman}}, \bibinfo{journal}{Phys. Rev. B}
  \textbf{\bibinfo{volume}{20}}, \bibinfo{pages}{624} (\bibinfo{year}{1979}).

\bibitem[{\citenamefont{Cui et~al.}(1998)\citenamefont{Cui, Ristein, and
  Ley}}]{CRL98}
\bibinfo{author}{\bibfnamefont{J.~B.} \bibnamefont{Cui}},
  \bibinfo{author}{\bibfnamefont{J.}~\bibnamefont{Ristein}}, \bibnamefont{and}
  \bibinfo{author}{\bibfnamefont{L.}~\bibnamefont{Ley}},
  \bibinfo{journal}{Phys. Rev. Lett.} \textbf{\bibinfo{volume}{81}},
  \bibinfo{pages}{429} (\bibinfo{year}{1998}).

\bibitem[{\citenamefont{Yamaguchi et~al.}(2009)\citenamefont{Yamaguchi,
  Masuzawa, Nozue, Kudo, Saito, Koe, Kudo, Yamada, Takakuwa, and
  Okano}}]{YMN09}
\bibinfo{author}{\bibfnamefont{H.}~\bibnamefont{Yamaguchi}},
  \bibinfo{author}{\bibfnamefont{T.}~\bibnamefont{Masuzawa}},
  \bibinfo{author}{\bibfnamefont{S.}~\bibnamefont{Nozue}},
  \bibinfo{author}{\bibfnamefont{Y.}~\bibnamefont{Kudo}},
  \bibinfo{author}{\bibfnamefont{I.}~\bibnamefont{Saito}},
  \bibinfo{author}{\bibfnamefont{J.}~\bibnamefont{Koe}},
  \bibinfo{author}{\bibfnamefont{M.}~\bibnamefont{Kudo}},
  \bibinfo{author}{\bibfnamefont{T.}~\bibnamefont{Yamada}},
  \bibinfo{author}{\bibfnamefont{Y.}~\bibnamefont{Takakuwa}}, \bibnamefont{and}
  \bibinfo{author}{\bibfnamefont{K.}~\bibnamefont{Okano}},
  \bibinfo{journal}{Phys. Rev. B} \textbf{\bibinfo{volume}{80}},
  \bibinfo{pages}{165321} (\bibinfo{year}{2009}).

\bibitem[{\citenamefont{Loh et~al.}(1999)\citenamefont{Loh, Sakaguchi, Gamo,
  Tagawa, Sugino, and Ando}}]{LSG99}
\bibinfo{author}{\bibfnamefont{K.~P.} \bibnamefont{Loh}},
  \bibinfo{author}{\bibfnamefont{I.}~\bibnamefont{Sakaguchi}},
  \bibinfo{author}{\bibfnamefont{M.~N.} \bibnamefont{Gamo}},
  \bibinfo{author}{\bibfnamefont{S.}~\bibnamefont{Tagawa}},
  \bibinfo{author}{\bibfnamefont{T.}~\bibnamefont{Sugino}}, \bibnamefont{and}
  \bibinfo{author}{\bibfnamefont{T.}~\bibnamefont{Ando}},
  \bibinfo{journal}{Appl. Phys. Lett.} \textbf{\bibinfo{volume}{74}},
  \bibinfo{pages}{28} (\bibinfo{year}{1999}).

\bibitem[{\citenamefont{Rohlfing et~al.}(2003)\citenamefont{Rohlfing, Wang,
  Kruger, and Pollmann}}]{RWK03}
\bibinfo{author}{\bibfnamefont{M.}~\bibnamefont{Rohlfing}},
  \bibinfo{author}{\bibfnamefont{N.-P.} \bibnamefont{Wang}},
  \bibinfo{author}{\bibfnamefont{P.}~\bibnamefont{Kruger}}, \bibnamefont{and}
  \bibinfo{author}{\bibfnamefont{J.}~\bibnamefont{Pollmann}},
  \bibinfo{journal}{Phys. Rev. Lett.} \textbf{\bibinfo{volume}{91}},
  \bibinfo{pages}{256802} (\bibinfo{year}{2003}).

\bibitem[{\citenamefont{Baumeier et~al.}(2007)\citenamefont{Baumeier, Kruger,
  and Pollmann}}]{BKP07}
\bibinfo{author}{\bibfnamefont{B.}~\bibnamefont{Baumeier}},
  \bibinfo{author}{\bibfnamefont{P.}~\bibnamefont{Kruger}}, \bibnamefont{and}
  \bibinfo{author}{\bibfnamefont{J.}~\bibnamefont{Pollmann}},
  \bibinfo{journal}{Phys. Rev. B} \textbf{\bibinfo{volume}{76}},
  \bibinfo{pages}{205404} (\bibinfo{year}{2007}).

\bibitem[{\citenamefont{McKenna and Shluger}(2008)}]{MS08}
\bibinfo{author}{\bibfnamefont{K.~P.} \bibnamefont{McKenna}} \bibnamefont{and}
  \bibinfo{author}{\bibfnamefont{A.~L.} \bibnamefont{Shluger}},
  \bibinfo{journal}{Nature Materials} \textbf{\bibinfo{volume}{7}},
  \bibinfo{pages}{859} (\bibinfo{year}{2008}).


\bibitem[{\citenamefont{Dinh et~al.}(1999)\citenamefont{Dinh, McLean, Schildbach,
  and Balooch}}]{DMS99}
\bibinfo{author}{\bibfnamefont{L.~N.} \bibnamefont{Dinh}},
  \bibinfo{author}{\bibfnamefont{W.} \bibnamefont{McLean}},
  \bibinfo{author}{\bibfnamefont{M.~A.} \bibnamefont{Schildbach}},
  \bibnamefont{and} \bibinfo{author}{\bibfnamefont{M.}~\bibnamefont{Balooch}},
  \bibinfo{journal}{Phys. Rev. B} \textbf{\bibinfo{volume}{59}},
  \bibinfo{pages}{15513} (\bibinfo{year}{1999}).

\bibitem[{\citenamefont{Geis et~al.}(2005)\citenamefont{Geis, Deneault, Krohn,
  Marchant, Lyszczarz, and Cooke}}]{GDK05}
\bibinfo{author}{\bibfnamefont{M.~W.} \bibnamefont{Geis}},
  \bibinfo{author}{\bibfnamefont{S.}~\bibnamefont{Deneault}},
  \bibinfo{author}{\bibfnamefont{K.~E.} \bibnamefont{Krohn}},
  \bibinfo{author}{\bibfnamefont{M.}~\bibnamefont{Marchant}},
  \bibinfo{author}{\bibfnamefont{T.~M.} \bibnamefont{Lyszczarz}},
  \bibnamefont{and} \bibinfo{author}{\bibfnamefont{D.~L.} \bibnamefont{Cooke}},
  \bibinfo{journal}{Appl. Phys. Lett.} \textbf{\bibinfo{volume}{87}},
  \bibinfo{pages}{192115} (\bibinfo{year}{2005}).

\bibitem[{\citenamefont{Mayer et~al.}(2006)\citenamefont{Mayer, Chung, Kumar,
  Weiss, Miskovsky, and Cutler}}]{MCK06}
\bibinfo{author}{\bibfnamefont{A.}~\bibnamefont{Mayer}},
  \bibinfo{author}{\bibfnamefont{M.~S.} \bibnamefont{Chung}},
  \bibinfo{author}{\bibfnamefont{N.}~\bibnamefont{Kumar}},
  \bibinfo{author}{\bibfnamefont{B.~L.} \bibnamefont{Weiss}},
  \bibinfo{author}{\bibfnamefont{N.~M.} \bibnamefont{Miskovsky}},
  \bibnamefont{and} \bibinfo{author}{\bibfnamefont{P.~H.}
  \bibnamefont{Cutler}}, \bibinfo{journal}{J. Vac. Sci. Technol. B}
  \textbf{\bibinfo{volume}{24}}, \bibinfo{pages}{1071} (\bibinfo{year}{2006}).

\bibitem[{\citenamefont{Biasini
  et~al.}(2005{\natexlab{a}})\citenamefont{Biasini, Gann, Yarmoff, Mills,
  Pfeiffer, West, Gao, and Williams}}]{BGY05}
\bibinfo{author}{\bibfnamefont{M.}~\bibnamefont{Biasini}},
  \bibinfo{author}{\bibfnamefont{R.~D.} \bibnamefont{Gann}},
  \bibinfo{author}{\bibfnamefont{J.~A.} \bibnamefont{Yarmoff}},
  \bibinfo{author}{\bibfnamefont{A.~P.} \bibnamefont{Mills}},
  \bibinfo{author}{\bibfnamefont{L.~N.} \bibnamefont{Pfeiffer}},
  \bibinfo{author}{\bibfnamefont{K.~W.} \bibnamefont{West}},
  \bibinfo{author}{\bibfnamefont{X.~P.~W.} \bibnamefont{Gao}},
  \bibnamefont{and} \bibinfo{author}{\bibfnamefont{B.~C.~D.}
  \bibnamefont{Williams}}, \bibinfo{journal}{Appl. Phys. Lett.}
  \textbf{\bibinfo{volume}{86}}, \bibinfo{pages}{162111}
  (\bibinfo{year}{2005}{\natexlab{a}}).

\bibitem[{\citenamefont{Biasini
  et~al.}(2005{\natexlab{b}})\citenamefont{Biasini, Gann, Pfeiffer, West, Gao,
  Williams, Yarmoff, and Jr.}}]{BGP05}
\bibinfo{author}{\bibfnamefont{M.}~\bibnamefont{Biasini}},
  \bibinfo{author}{\bibfnamefont{R.~D.} \bibnamefont{Gann}},
  \bibinfo{author}{\bibfnamefont{L.~N.} \bibnamefont{Pfeiffer}},
  \bibinfo{author}{\bibfnamefont{K.~W.} \bibnamefont{West}},
  \bibinfo{author}{\bibfnamefont{X.~P.~W.} \bibnamefont{Gao}},
  \bibinfo{author}{\bibfnamefont{B.~C.~D.} \bibnamefont{Williams}},
  \bibinfo{author}{\bibfnamefont{J.~A.} \bibnamefont{Yarmoff}},
  \bibnamefont{and} \bibinfo{author}{\bibfnamefont{A.~P.~M.}
  \bibnamefont{Jr.}}, \bibinfo{journal}{Eur. Phys. J. B}
  \textbf{\bibinfo{volume}{47}}, \bibinfo{pages}{305}
  (\bibinfo{year}{2005}{\natexlab{b}}).

\bibitem[{\citenamefont{Yamamoto et~al.}(2007)\citenamefont{Yamamoto, Yamamoto,
  Kuwahara, Sakai, Morino, Tamagaki, Mano, Utsu, Okumi, Nakanishi
  et~al.}}]{YYK07}
\bibinfo{author}{\bibfnamefont{N.}~\bibnamefont{Yamamoto}},
  \bibinfo{author}{\bibfnamefont{M.}~\bibnamefont{Yamamoto}},
  \bibinfo{author}{\bibfnamefont{K.}~\bibnamefont{Kuwahara}},
  \bibinfo{author}{\bibfnamefont{R.}~\bibnamefont{Sakai}},
  \bibinfo{author}{\bibfnamefont{T.}~\bibnamefont{Morino}},
  \bibinfo{author}{\bibfnamefont{K.}~\bibnamefont{Tamagaki}},
  \bibinfo{author}{\bibfnamefont{A.}~\bibnamefont{Mano}},
  \bibinfo{author}{\bibfnamefont{A.}~\bibnamefont{Utsu}},
  \bibinfo{author}{\bibfnamefont{S.}~\bibnamefont{Okumi}},
  \bibinfo{author}{\bibfnamefont{T.}~\bibnamefont{Nakanishi}},
  \bibnamefont{et~al.}, \bibinfo{journal}{J. Appl. Phys.}
  \textbf{\bibinfo{volume}{102}}, \bibinfo{pages}{024904}
  (\bibinfo{year}{2007}).

\bibitem[{\citenamefont{Desjonqueres and Spanjaard}(1996)}]{Spanjaard96}
\bibinfo{author}{\bibfnamefont{M.-C.} \bibnamefont{Desjonqueres}}
  \bibnamefont{and}
  \bibinfo{author}{\bibfnamefont{D.}~\bibnamefont{Spanjaard}},
  \emph{\bibinfo{title}{Concepts of surface physics}}
  (\bibinfo{publisher}{Springer Verlag}, \bibinfo{address}{Berlin},
  \bibinfo{year}{1996}).

\bibitem[{\citenamefont{Lennard-Jones and Strachan}(1935)}]{LJS35}
\bibinfo{author}{\bibfnamefont{J.~E.} \bibnamefont{Lennard-Jones}}
  \bibnamefont{and} \bibinfo{author}{\bibfnamefont{C.}~\bibnamefont{Strachan}},
  \bibinfo{journal}{Proc. Roy. Soc. London, Ser. A}
  \textbf{\bibinfo{volume}{150}}, \bibinfo{pages}{442} (\bibinfo{year}{1935}).

\bibitem[{\citenamefont{Strachan}(1935)}]{Strachan35}
\bibinfo{author}{\bibfnamefont{C.}~\bibnamefont{Strachan}},
  \bibinfo{journal}{Proc. Roy. Soc. London, Ser. A}
  \textbf{\bibinfo{volume}{150}}, \bibinfo{pages}{456} (\bibinfo{year}{1935}).

\bibitem[{\citenamefont{Lennard-Jones and Devonshire}(1936)}]{LJD36}
\bibinfo{author}{\bibfnamefont{J.~E.} \bibnamefont{Lennard-Jones}}
  \bibnamefont{and} \bibinfo{author}{\bibfnamefont{A.~F.}
  \bibnamefont{Devonshire}}, \bibinfo{journal}{Proc. Roy. Soc. London, Ser. A}
  \textbf{\bibinfo{volume}{156}}, \bibinfo{pages}{6} (\bibinfo{year}{1936}).

\bibitem[{\citenamefont{Bendow and Ying}(1973)}]{BY73}
\bibinfo{author}{\bibfnamefont{B.}~\bibnamefont{Bendow}} \bibnamefont{and}
  \bibinfo{author}{\bibfnamefont{S.-C.} \bibnamefont{Ying}},
  \bibinfo{journal}{Phys. Rev. B} \textbf{\bibinfo{volume}{7}},
  \bibinfo{pages}{622} (\bibinfo{year}{1973}).

\bibitem[{\citenamefont{Brenig}(1982)}]{Brenig82}
\bibinfo{author}{\bibfnamefont{W.}~\bibnamefont{Brenig}}, \bibinfo{journal}{Z.
  Phys. B} \textbf{\bibinfo{volume}{48}}, \bibinfo{pages}{127}
  (\bibinfo{year}{1982}).

\bibitem[{\citenamefont{Gortel et~al.}(1980{\natexlab{a}})\citenamefont{Gortel,
  Kreuzer, and Spaner}}]{GKS80}
\bibinfo{author}{\bibfnamefont{Z.~W.} \bibnamefont{Gortel}},
  \bibinfo{author}{\bibfnamefont{H.~J.} \bibnamefont{Kreuzer}},
  \bibnamefont{and} \bibinfo{author}{\bibfnamefont{D.}~\bibnamefont{Spaner}},
  \bibinfo{journal}{J. Chem. Phys,} \textbf{\bibinfo{volume}{72}},
  \bibinfo{pages}{234} (\bibinfo{year}{1980}{\natexlab{a}}).

\bibitem[{\citenamefont{Gortel et~al.}(1980{\natexlab{b}})\citenamefont{Gortel,
  Kreuzer, and Teshima}}]{GKT80a}
\bibinfo{author}{\bibfnamefont{Z.~W.} \bibnamefont{Gortel}},
  \bibinfo{author}{\bibfnamefont{H.~J.} \bibnamefont{Kreuzer}},
  \bibnamefont{and} \bibinfo{author}{\bibfnamefont{R.}~\bibnamefont{Teshima}},
  \bibinfo{journal}{Phys. Rev. B} \textbf{\bibinfo{volume}{22}},
  \bibinfo{pages}{5655} (\bibinfo{year}{1980}{\natexlab{b}}).

\bibitem[{\citenamefont{Gortel et~al.}(1980{\natexlab{c}})\citenamefont{Gortel,
  Kreuzer, and Teshima}}]{GKT80b}
\bibinfo{author}{\bibfnamefont{Z.~W.} \bibnamefont{Gortel}},
  \bibinfo{author}{\bibfnamefont{H.~J.} \bibnamefont{Kreuzer}},
  \bibnamefont{and} \bibinfo{author}{\bibfnamefont{R.}~\bibnamefont{Teshima}},
  \bibinfo{journal}{Phys. Rev. B} \textbf{\bibinfo{volume}{22}},
  \bibinfo{pages}{512} (\bibinfo{year}{1980}{\natexlab{c}}).

\bibitem[{\citenamefont{Kreuzer and Gortel}(1986)}]{KG86}
\bibinfo{author}{\bibfnamefont{H.~J.} \bibnamefont{Kreuzer}} \bibnamefont{and}
  \bibinfo{author}{\bibfnamefont{Z.~W.} \bibnamefont{Gortel}},
  \emph{\bibinfo{title}{Physisorption Kinetics}} (\bibinfo{publisher}{Springer
  Verlag}, \bibinfo{address}{Berlin}, \bibinfo{year}{1986}).

\bibitem[{\citenamefont{Brenig}(1987)}]{Brenig87}
\bibinfo{author}{\bibfnamefont{W.}~\bibnamefont{Brenig}},
  \bibinfo{journal}{Physica Scripta} \textbf{\bibinfo{volume}{35}},
  \bibinfo{pages}{329} (\bibinfo{year}{1987}).

\bibitem[{\citenamefont{Bronold et~al.}(2008)\citenamefont{Bronold, Fehske,
  Kersten, and Deutsch}}]{BFKD08}
\bibinfo{author}{\bibfnamefont{F.~X.} \bibnamefont{Bronold}},
  \bibinfo{author}{\bibfnamefont{H.}~\bibnamefont{Fehske}},
  \bibinfo{author}{\bibfnamefont{H.}~\bibnamefont{Kersten}}, \bibnamefont{and}
  \bibinfo{author}{\bibfnamefont{H.}~\bibnamefont{Deutsch}},
  \bibinfo{journal}{Phys. Rev. Lett.} \textbf{\bibinfo{volume}{101}},
  \bibinfo{pages}{175002} (\bibinfo{year}{2008}).

\bibitem[{\citenamefont{Bronold et~al.}(2009)\citenamefont{Bronold, Deutsch,
  and Fehske}}]{BDF09}
\bibinfo{author}{\bibfnamefont{F.~X.} \bibnamefont{Bronold}},
  \bibinfo{author}{\bibfnamefont{H.}~\bibnamefont{Deutsch}}, \bibnamefont{and}
  \bibinfo{author}{\bibfnamefont{H.}~\bibnamefont{Fehske}},
  \bibinfo{journal}{Eur. Phys. J. D} \textbf{\bibinfo{volume}{54}},
  \bibinfo{pages}{519} (\bibinfo{year}{2009}).

\bibitem[{\citenamefont{Whipple}(1981)}]{Whipple81}
\bibinfo{author}{\bibfnamefont{E.~C.} \bibnamefont{Whipple}},
  \bibinfo{journal}{Rep. Prog. Phys.} \textbf{\bibinfo{volume}{44}},
  \bibinfo{pages}{1197} (\bibinfo{year}{1981}).

\bibitem[{\citenamefont{Draine and Sutin}(1987)}]{DS87}
\bibinfo{author}{\bibfnamefont{B.~T.} \bibnamefont{Draine}} \bibnamefont{and}
  \bibinfo{author}{\bibfnamefont{B.}~\bibnamefont{Sutin}},
  \bibinfo{journal}{The Astrophysical Journal} \textbf{\bibinfo{volume}{320}},
  \bibinfo{pages}{803} (\bibinfo{year}{1987}).

\bibitem[{\citenamefont{Mann}(2008)}]{Mann08}
\bibinfo{author}{\bibfnamefont{I.}~\bibnamefont{Mann}},
  \bibinfo{journal}{Advances in Space Research} \textbf{\bibinfo{volume}{41}},
  \bibinfo{pages}{160} (\bibinfo{year}{2008}).

\bibitem[{\citenamefont{Rapp and Luebken}(2001)}]{RL01}
\bibinfo{author}{\bibfnamefont{M.}~\bibnamefont{Rapp}} \bibnamefont{and}
  \bibinfo{author}{\bibfnamefont{F.-J.} \bibnamefont{Luebken}},
  \bibinfo{journal}{J. Atmospheric and solar-terrestrial physics}
  \textbf{\bibinfo{volume}{63}}, \bibinfo{pages}{759} (\bibinfo{year}{2001}).

\bibitem[{\citenamefont{Fortov et~al.}(2005)\citenamefont{Fortov, Ivlev,
  Khrapak, Khrapak, and Morfill}}]{FIK05}
\bibinfo{author}{\bibfnamefont{V.~E.} \bibnamefont{Fortov}},
  \bibinfo{author}{\bibfnamefont{A.~V.} \bibnamefont{Ivlev}},
  \bibinfo{author}{\bibfnamefont{S.~A.} \bibnamefont{Khrapak}},
  \bibinfo{author}{\bibfnamefont{A.~G.} \bibnamefont{Khrapak}},
  \bibnamefont{and} \bibinfo{author}{\bibfnamefont{G.~E.}
  \bibnamefont{Morfill}}, \bibinfo{journal}{Physics Reports}
  \textbf{\bibinfo{volume}{421}}, \bibinfo{pages}{1} (\bibinfo{year}{2005}).

\bibitem[{\citenamefont{Ishihara}(2007)}]{Ishihara07}
\bibinfo{author}{\bibfnamefont{O.}~\bibnamefont{Ishihara}},
  \bibinfo{journal}{J. Phys. D: Appl. Phys} \textbf{\bibinfo{volume}{40}},
  \bibinfo{pages}{R121} (\bibinfo{year}{2007}).

\bibitem[{\citenamefont{Golubovskii et~al.}(2002)\citenamefont{Golubovskii,
  Maiorov, Behnke, and Behnke}}]{GMB02}
\bibinfo{author}{\bibfnamefont{Y.~B.} \bibnamefont{Golubovskii}},
  \bibinfo{author}{\bibfnamefont{V.~A.} \bibnamefont{Maiorov}},
  \bibinfo{author}{\bibfnamefont{J.}~\bibnamefont{Behnke}}, \bibnamefont{and}
  \bibinfo{author}{\bibfnamefont{J.~F.} \bibnamefont{Behnke}},
  \bibinfo{journal}{J. Phys. D: Appl. Phys} \textbf{\bibinfo{volume}{35}},
  \bibinfo{pages}{751} (\bibinfo{year}{2002}).

\bibitem[{\citenamefont{Kogelschatz}(2003)}]{Kogelschatz03}
\bibinfo{author}{\bibfnamefont{U.}~\bibnamefont{Kogelschatz}},
  \bibinfo{journal}{Plasma Chemistry and Plasma Processing}
  \textbf{\bibinfo{volume}{23}}, \bibinfo{pages}{1} (\bibinfo{year}{2003}).

\bibitem[{\citenamefont{Li et~al.}(2004)\citenamefont{Li, Li, Zhan, and
  Xu}}]{KCO04}
\bibinfo{author}{\bibfnamefont{M.}~\bibnamefont{Li}},
  \bibinfo{author}{\bibfnamefont{C.}~\bibnamefont{Li}},
  \bibinfo{author}{\bibfnamefont{H.}~\bibnamefont{Zhan}}, \bibnamefont{and}
  \bibinfo{author}{\bibfnamefont{J.}~\bibnamefont{Xu}},
  \bibinfo{journal}{Proceedings of the XV International Conference on Gas
  Discharges and their Applications}  (\bibinfo{year}{2004}).

\bibitem[{\citenamefont{Stollenwerk et~al.}(2006)\citenamefont{Stollenwerk,
  Amiranashvili, Boeuf, and Purwins}}]{SAB06}
\bibinfo{author}{\bibfnamefont{L.}~\bibnamefont{Stollenwerk}},
  \bibinfo{author}{\bibfnamefont{S.}~\bibnamefont{Amiranashvili}},
  \bibinfo{author}{\bibfnamefont{J.-P.} \bibnamefont{Boeuf}}, \bibnamefont{and}
  \bibinfo{author}{\bibfnamefont{H.-G.} \bibnamefont{Purwins}},
  \bibinfo{journal}{Phys. Rev. Lett.} \textbf{\bibinfo{volume}{96}},
  \bibinfo{pages}{255001} (\bibinfo{year}{2006}).

\bibitem[{\citenamefont{Stollenwerk et~al.}(2007)\citenamefont{Stollenwerk,
  Laven, and Purwins}}]{SLP07}
\bibinfo{author}{\bibfnamefont{L.}~\bibnamefont{Stollenwerk}},
  \bibinfo{author}{\bibfnamefont{J.~G.} \bibnamefont{Laven}}, \bibnamefont{and}
  \bibinfo{author}{\bibfnamefont{H.-G.} \bibnamefont{Purwins}},
  \bibinfo{journal}{Phys. Rev. Lett.} \textbf{\bibinfo{volume}{98}},
  \bibinfo{pages}{255001} (\bibinfo{year}{2007}).

\bibitem[{\citenamefont{Li et~al.}(2008)\citenamefont{Li, Li, Zhan, and
  Xu}}]{LLZ08}
\bibinfo{author}{\bibfnamefont{M.}~\bibnamefont{Li}},
  \bibinfo{author}{\bibfnamefont{C.}~\bibnamefont{Li}},
  \bibinfo{author}{\bibfnamefont{H.}~\bibnamefont{Zhan}}, \bibnamefont{and}
  \bibinfo{author}{\bibfnamefont{J.}~\bibnamefont{Xu}}, \bibinfo{journal}{Appl.
  Phys. Lett.} \textbf{\bibinfo{volume}{92}}, \bibinfo{pages}{031503}
  (\bibinfo{year}{2008}).

\bibitem[{\citenamefont{Kus\u{c}er}(1978)}]{Kuscer78}
\bibinfo{author}{\bibfnamefont{I.}~\bibnamefont{Kus\u{c}er}}, in
  \emph{\bibinfo{booktitle}{Fundamental problems in statistical physics IV}},
  edited by \bibinfo{editor}{\bibfnamefont{E.~G.~D.} \bibnamefont{Cohen}}
  \bibnamefont{and} \bibinfo{editor}{\bibfnamefont{W.}~\bibnamefont{Fiszdon}}
  (\bibinfo{publisher}{Ossolineum}, \bibinfo{address}{Warsaw},
  \bibinfo{year}{1978}), p. \bibinfo{pages}{441}.

\bibitem[{\citenamefont{Fan and Manson}(2009{\natexlab{a}})}]{FM09a}
\bibinfo{author}{\bibfnamefont{G.}~\bibnamefont{Fan}} \bibnamefont{and}
  \bibinfo{author}{\bibfnamefont{J.~R.} \bibnamefont{Manson}},
  \bibinfo{journal}{Phys. Rev. B} \textbf{\bibinfo{volume}{79}},
  \bibinfo{pages}{045424} (\bibinfo{year}{2009}{\natexlab{a}}).

\bibitem[{\citenamefont{Fan and Manson}(2009{\natexlab{b}})}]{FM09b}
\bibinfo{author}{\bibfnamefont{G.}~\bibnamefont{Fan}} \bibnamefont{and}
  \bibinfo{author}{\bibfnamefont{J.~R.} \bibnamefont{Manson}},
  \bibinfo{journal}{J. Chem. Phys.} \textbf{\bibinfo{volume}{130}},
  \bibinfo{pages}{064703} (\bibinfo{year}{2009}{\natexlab{b}}).

\bibitem[{\citenamefont{Neilson et~al.}(1986)\citenamefont{Neilson, Nieminen,
  and Szyma\'nski}}]{NNS86}
\bibinfo{author}{\bibfnamefont{D.}~\bibnamefont{Neilson}},
  \bibinfo{author}{\bibfnamefont{R.~M.} \bibnamefont{Nieminen}},
  \bibnamefont{and}
  \bibinfo{author}{\bibfnamefont{J.}~\bibnamefont{Szyma\'nski}},
  \bibinfo{journal}{Phys. Rev. B} \textbf{\bibinfo{volume}{33}},
  \bibinfo{pages}{1567} (\bibinfo{year}{1986}).

\bibitem[{\citenamefont{Walker et~al.}(1992)\citenamefont{Walker, Jensen,
  Szyma\'nski, and Neilson}}]{WJS92}
\bibinfo{author}{\bibfnamefont{A.~B.} \bibnamefont{Walker}},
  \bibinfo{author}{\bibfnamefont{K.~O.} \bibnamefont{Jensen}},
  \bibinfo{author}{\bibfnamefont{J.}~\bibnamefont{Szyma\'nski}},
  \bibnamefont{and} \bibinfo{author}{\bibfnamefont{D.}~\bibnamefont{Neilson}},
  \bibinfo{journal}{Phys. Rev. B} \textbf{\bibinfo{volume}{46}},
  \bibinfo{pages}{1687} (\bibinfo{year}{1992}).

\bibitem[{\citenamefont{Ray and Mahan}(1972)}]{RM72}
\bibinfo{author}{\bibfnamefont{R.}~\bibnamefont{Ray}} \bibnamefont{and}
  \bibinfo{author}{\bibfnamefont{G.~D.} \bibnamefont{Mahan}},
  \bibinfo{journal}{Phys. Lett.} \textbf{\bibinfo{volume}{42A}},
  \bibinfo{pages}{301} (\bibinfo{year}{1972}).

\bibitem[{\citenamefont{Evans and Mills}(1973)}]{EM73}
\bibinfo{author}{\bibfnamefont{E.}~\bibnamefont{Evans}} \bibnamefont{and}
  \bibinfo{author}{\bibfnamefont{D.~L.} \bibnamefont{Mills}},
  \bibinfo{journal}{Phys. Rev. B} \textbf{\bibinfo{volume}{8}},
  \bibinfo{pages}{4004} (\bibinfo{year}{1973}).

\bibitem[{\citenamefont{Gumhalter}(1996)}]{Gumhalter96}
\bibinfo{author}{\bibfnamefont{B.}~\bibnamefont{Gumhalter}},
  \bibinfo{journal}{Surface science} \textbf{\bibinfo{volume}{347}},
  \bibinfo{pages}{237} (\bibinfo{year}{1996}).

\bibitem[{\citenamefont{Heinisch}(2009)}]{Rafael09}
\bibinfo{author}{\bibfnamefont{R.~L.} \bibnamefont{Heinisch}},
  \emph{\bibinfo{title}{Trapping and detrapping of charged particles at
  surfaces}}, \bibinfo{howpublished}{Diploma thesis (Universit\"at Greifswald)}
  (\bibinfo{year}{2009}).

\bibitem[{\citenamefont{Klemens}(2001)}]{Klemens01}
\bibinfo{author}{\bibfnamefont{P.~G.} \bibnamefont{Klemens}},
  \bibinfo{journal}{Int. J. of Thermophysics} \textbf{\bibinfo{volume}{22}},
  \bibinfo{pages}{265} (\bibinfo{year}{2001}).

\bibitem[{\citenamefont{Whittaker and Watson}(1927)}]{WW27}
\bibinfo{author}{\bibfnamefont{E.~T.} \bibnamefont{Whittaker}}
  \bibnamefont{and} \bibinfo{author}{\bibfnamefont{G.~N.}
  \bibnamefont{Watson}}, \emph{\bibinfo{title}{A course of modern analysis}}
  (\bibinfo{publisher}{Cambridge University Press}, \bibinfo{year}{1927}).

\bibitem[{\citenamefont{Magnus and Oberhettinger}(1948)}]{MO48}
\bibinfo{author}{\bibfnamefont{W.}~\bibnamefont{Magnus}} \bibnamefont{and}
  \bibinfo{author}{\bibfnamefont{F.}~\bibnamefont{Oberhettinger}},
  \emph{\bibinfo{title}{Formeln und S\"atze f\"ur die speziellen Funktionen der
  mathematischen Physik}} (\bibinfo{publisher}{Springer},
  \bibinfo{year}{1948}).

\bibitem[{\citenamefont{Abramowitz and Stegun}(1973)}]{AS73}
\bibinfo{editor}{\bibfnamefont{M.}~\bibnamefont{Abramowitz}} \bibnamefont{and}
  \bibinfo{editor}{\bibfnamefont{I.~A.} \bibnamefont{Stegun}}, eds.,
  \emph{\bibinfo{title}{Handbook of mathematical functions}}
  (\bibinfo{publisher}{Dover Publications, Inc.}, \bibinfo{address}{New York},
  \bibinfo{year}{1973}).

\bibitem[{\citenamefont{Olver}(1974)}]{Olver74}
\bibinfo{author}{\bibfnamefont{F.~W.~J.} \bibnamefont{Olver}},
  \emph{\bibinfo{title}{Aysmptotics and special functions}}
  (\bibinfo{publisher}{Academic Press}, \bibinfo{address}{New York},
  \bibinfo{year}{1974}).

\end{thebibliography}
\end{document}